\documentclass[%
reprint,
 amsmath,amssymb,
 aps
]{revtex4-2}

\usepackage{graphicx} 
\usepackage{subfigure}
\usepackage{array}
\usepackage{makecell}
\usepackage{comment}
\usepackage{dcolumn}
\usepackage{bm}
\usepackage{stmaryrd}
\usepackage{algpseudocode}
\usepackage{algorithm}
\usepackage{multirow}
\usepackage{acronym}
\usepackage{xcolor}
\usepackage{amsmath}
\usepackage{mathtools}

\begin{document}

\title{
Flow of temporal network properties under local aggregation and time shuffling: \\
a tool for characterizing, comparing and classifying temporal networks
}

\author{Didier Le Bail}
\affiliation{Aix Marseille Univ, Universit\'e de Toulon, CNRS, CPT, Marseille, France}
\author{Mathieu G\'enois}%
\affiliation{Aix Marseille Univ, Universit\'e de Toulon, CNRS, CPT, Marseille, France}
\author{Alain Barrat}
\affiliation{Aix Marseille Univ, Universit\'e de Toulon, CNRS, CPT, Marseille, France}
\email{alain.barrat@cpt.univ-mrs.fr}

\date{\today}

\begin{abstract}
Although many tools have been developed and employed to characterize temporal networks, the issue of how to
compare them remains largely open.
It depends indeed on what features are considered as relevant, and on the way the differences in these features are quantified.
In this paper, we propose to characterize temporal networks through their behaviour under general transformations that are local in time:
(i) a local time shuffling, which destroys correlations at time scales smaller than a given scale $b$, while preserving large time scales,
and (ii) a local temporal aggregation on time windows of length $n$.
By varying $b$ and $n$, we obtain a flow of temporal networks, and flows of observable values, which encode the phenomenology of the temporal network on multiple time scales.
We use a symbolic approach to summarize these flows into labels (strings of characters) describing their trends. These labels can then be used to compare temporal networks, validate models, or identify groups of networks with similar labels. Our procedure can 
be applied to any temporal network and 
with an arbitrary set of observables, and we illustrate it on an  ensemble of data sets describing face-to-face interactions in various contexts, including both empirical and synthetic data.
\end{abstract}

\maketitle



\section{Introduction}

A large variety of natural and artificial systems can be described as networks, in which nodes represent the elements of the system and links represent their interactions. The corresponding data  
are increasingly available with temporal resolution, which has led to the development of the study of temporal networks, in which each link can be alternatively active or inactive \cite{holme2015modern,holme2012temporal}. A growing literature 
deals with the development of analysis tools for these complex objects, and of models to describe them. 
In this context, here we are interested in the issue of characterizing, comparing and classifying temporal networks or ensembles of temporal networks: being able to compare quantitatively networks is indeed needed for instance to detect differences between data sets of an a priori similar nature, or to validate models and quantify how well they represent data. This issue is also crucial for static networks, for which a large set of comparison methods has been developed \cite{berlingerio2013network,bagrow2019information,tantardini2019comparing,hartle2020network}, but 
the case of temporal networks has barely been considered \cite{holme2012temporal,tu2018network,zhan2021measuring}. 
When are two temporal networks equivalent or what is their degree of similarity? Depending on the properties that are considered as most relevant, several methods can be devised to answer these questions: if two temporal networks share such properties, they are then considered as equivalent.
For instance one can generalize notions of distance between static networks, using distributions of path lengths \cite{zhan2021measuring} or statistics of motifs \cite{tu2018network}. One can also consider a temporal network as a multi-dimensional time series and characterize these series e.g. by their Fourier transform or other tools \cite{andres2023detecting,sikdar2016time,nie2023topological}.

As in the case of static networks, another approach lies in adopting a statistical point of view, by considering a set of variables (observables) that can be sampled from the network, and their resulting distributions \cite{berlingerio2013network}: these distributions are then considered as the set of relevant properties able to characterize a data set or a model, and the similarity of these distributions is seen as corresponding to the similarity of the data or as the ability of the model to represent the data \cite{perra2012activity,barrat2013temporal,vestergaard2014memory,zhao2011social,Laurent_2015,longa2022neighbourhood,le2023modeling}.
This statistical point of view comes from the underlying hypotheses that the empirical temporal network under study results from some stochastic process (often implicitly approximated as stationary), and that two sets of similar distributions of observables comes from similar stochastic processes.
In temporal networks for instance, observables commonly used to this purpose include
the duration of an interaction and the time elapsed between two consecutive interactions \cite{tang2009temporal,nicosia2013graph,Karsai2012,cattuto2010dynamics,vestergaard2014memory}. This approach has several limitations: (1) the sampling of observables is usually performed at the temporal resolution given by the data at hand, while network dynamics exist typically at many timescales, and moreover different temporal networks might be collected with different time resolutions \cite{toth2015role,mastrandrea2015contact,genois2022combining,sapiezynski2019interaction}; the 
choice of an adequate resolution is non trivial as every scale of observation can potentially give different information about a social system \cite{krings2012effects,sulo2010meaningful};
(2) moreover, sharing the same distributions does not mean sharing the same correlations, so that sampling observables and obtaining their distributions might not be enough to characterize a temporal network \cite{Karsai2012};
(3) distributions are in general difficult to compare, especially if they present broad functional shapes \cite{clauset2009power,voitalov2019scale}.

In this article, we present a methodology to go beyond these limitations. To this aim, we propose a labelling procedure of temporal networks, based on their transformation under reshuffling at a certain scale and temporal aggregation at another scale. 
By varying the scales of reshuffling and aggregation, we indeed create a two-dimensional flow for the temporal network and for any attached scalar observable of interest. This allows to go beyond limitation (1) by including information on all time scales in the resulting label. Moreover, by using a reshuffling procedure at a given scale, we remove short scale temporal correlations while preserving long range ones (intuitively, this amounts to apply a low-pass filter acting on the time correlation function). By varying this scale, we include information about not only distributions of observables but about correlations in the temporal network, going beyond the limitation (2). As comparing flows of observables is not an easy task, we simply summarize each flow of a scalar observable under a one-parameter transformation by the sign sequence of its derivative, which encodes the shape of the corresponding curve. As we have a two-dimensional flow, we have thus two varying parameters (the scales of time shuffling and of time aggregation) and hence two sets of sign sequences that we combine to form two sentences. Each temporal network is thus finally labelled by strings (two for each observable of interest), which can then be compared. 
We can now define an equivalence between temporal networks as the fact that they share the same label, and we can define a distance between two temporal networks through a distance between labels (e.g., the edit distance). Note that this procedure is limited to scalar observables (having a single realization per temporal network), while the focus is often on distributions, e.g. of interaction durations. As in \cite{berlingerio2013network}, we will thus reduce each distribution of interest to its moments.

In the following, we describe step by step the characterization procedure and how to go from a temporal network to a label.
We  then introduce tools to analyze those labels, use them to compare temporal networks or groups of temporal networks.
Finally, we  illustrate our procedure and tools of analysis with the study of 27 temporal networks through the flows of 43 different observables \cite{genois2022combining,
Gemmetto2014,
10.1371/journal.pone.0023176,
10.1371/journal.pone.0107878,
mastrandrea2015contact,
10.1371/journal.pone.0073970,
kiti2016quantifying,
gelardi2020measuring,
NWS:9950811,
Genois2018,
Laurent_2015,
le2023modeling,
Longa2022,
perra2012activity,
Karsai2012,
vicsek1995novel}. 
We note that our procedure can be applied to any temporal networks as long as they are defined in discrete time.
Moreover, it can be tailored to any specific field through an adequate choice of observables relevant to that field. 
Here, we will focus for illustration purposes on temporal networks describing face-to-face interactions between individuals in various social contexts. On the one hand indeed, a number of
data sets describing such contacts in various contexts are publicly available \cite{toth2015role,sociopatterns}. 
On the other hand, a number of temporal network models have been proposed to mimic the mechanisms at play in social networks and the resulting phenomenology, and we will also consider several of those \cite{perra2012activity,Laurent_2015,le2023modeling,vicsek1995novel}.

\section{From a temporal network to a label}

The first step in labelling a temporal network (TN) is to create the transformation flow under removal of short-range time correlations and time aggregation. Each of these transformations is characterized by a single integer parameter, that we will denote by $b$ for the reshuffling procedure and $n$ for the time aggregation in the remainder of the paper.

\begin{figure}
    \centering
    \includegraphics[width=\columnwidth]{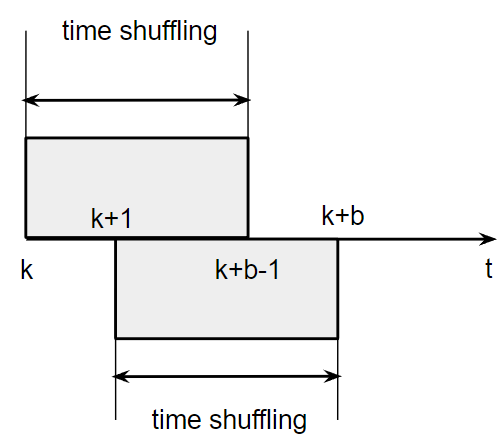}
    \caption{\label{fig:0}\textbf{Illustration of sliding time shuffling.}
    $b$ is the integer parameter of the sliding time shuffling, called its range.
    $k$ denotes a timestamp in the TN timeline.
    Time shuffling is done iteratively following this timeline, so the transformation is not strictly local. However, the distance travelled by a timestamp follows an exponential law, with average $2b$.
    This removes any correlation on time scales shorted than $2b$ but preserves correlations on larger time scales.
    }
\end{figure}

Let us consider a temporal network ${\cal G}$ in discrete time $t=0,\cdots,T-1$. We denote by $G(t)$ 
the snapshot at time $t$, which is formed by all active edges at $t$.
We first proceed with the time shuffling (filtering) transformation using a sliding time window, as illustrated in Fig. \ref{fig:0}, i.e., we apply a sliding time shuffling (STS).
Specifically, to apply a STS of range $b$ to a TN, we browse its timeline one time step after the other.
At each time step $k$, we reshuffle at random the $b$ following temporal snapshots $G(k),\cdots,G(k+b-1)$ \cite{gauvin2022randomized}. The procedure is then iterated at time $k+1$
\footnote{Note that instead of using a sliding procedure, we could divide the TN timeline into disjoint blocks of length $b$ and shuffle each block separately. As discussed in Appendix \ref{sec:AppendixAA}, this turns out to introduce spurious deterministic noise and oscillations in the values of the observables.}.
Successively, we aggregate the temporal network obtained after reshuffling at level $n$ using sliding windows as well: each snapshot $G(t)$ is replaced by
$G^{(n)}(t)$ resulting from the aggregation
of snapshots $G(t),G(t+1),\cdots,G(t+n-1)$.

\begin{figure}
\subfigure[conf16]{\includegraphics[width=\columnwidth]{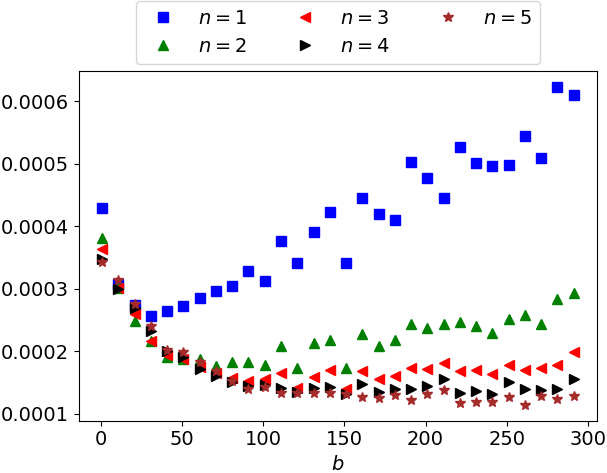}}
\subfigure[ADM18conf16]{\includegraphics[width=\columnwidth]{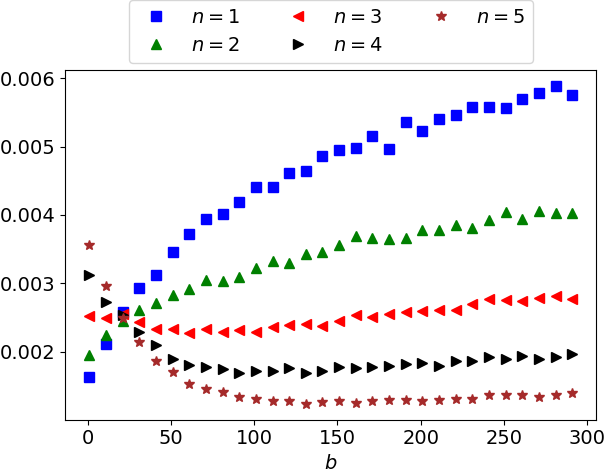}}
    \caption{\label{fig:2}\textbf{Bi-dimensional flows of a specific observable ``3-NCTN\_error''
    in two temporal networks.} (See Appendix \ref{sec:app_nomenclature} for the explanation of the nomenclature of observables.)
    Here, bi-dimensional flows are presented as collections of partial flows versus $b$, for various values of $n$ (cf text).
    (a) data set ``conf16'' (describing contacts in a scientific conference \cite{genois2022combining}), (b) data set ``ADM18conf16'' (model of temporal network from the ADM class \cite{le2023modeling})
    }
\end{figure}

We denote by TN$(n,b)$ the temporal network obtained after a STS of range $b$ followed by a time aggregation at level $n$. TN$(n=1,b=1)$ corresponds to the original data set.
Then, for each observable $\mathcal{O}$, we can compute its realization $\mathcal{O}(n,b)$ in the transformed network TN$(n,b)$. Note that, 
to reduce the noise originating from the random shufflings, we average observable values over ten independent STS realizations.

Collecting values $\mathcal{O}(n,b)$ for various $n$ and $b$ results in a bi-dimensional flow $(n,b)\mapsto\mathcal{O}(n,b)$ for each observable $\mathcal{O}$.
This global flow can be viewed as a collection indexed by $n$ (resp. indexed by $b$) of one dimensional flows versus $b$ (resp. versus $n$), see Fig. \ref{fig:2} for examples.
For a given value of $b$, the partial flow versus $n$ is written as $\mathcal{O}(.,b)$. Symmetrically, the partial flow versus $b$, at fixed $n$, is written as $\mathcal{O}(n,.)$.

Each partial flow is a one dimensional curve, describing how the associated observable is changing under the associated transformation:
$\mathcal{O}(n,.)$ tells us how $\mathcal{O}$ is changing under sliding time shuffling with varying $b$, while $\mathcal{O}(.,b)$ determines how $\mathcal{O}$ is changing under sliding time aggregation of varying length $n$. We hypothesize that important information on the temporal network structure and correlations is contained in the shape of these flows. Therefore, we do not focus on the specific values of the observables but on the trends of their evolution with $n$ and $b$. We thus need to extract such information from the derivatives of 
$\mathcal{O}(.,b)$ with respect to $n$ and of 
$\mathcal{O}(n,.)$ with respect to $b$. Moreover, in order to focus on trends, we do not collect derivatives in each point of the curves but in the observed succession of trends, such as, e.g., ``the curve first increases then decreases''.

To extract automatically this information, we determine the denoised sign sequence of the partial flows derivatives using an artificial neural network that we build for this purpose (see Appendix \ref{sec:AppendixF} for the description of the neural network and of the training procedure). 
The resulting sign sequence describing each partial flow is encoded as a word made up with letters ``+'' (for increasing parts of the curve), ``$-$'' and ``0'' (for decreasing and flat parts of the curve). Note that consecutive letters in a word are always different because, as discussed above, we focus on the successive trends 
(the successive signs taken by the derivative), rather than on its sign at every point of the flow.
As a result, each partial flow versus $n$ yields a word ``versus $n$'', i.e., describing its behaviour versus $n$, and we thus obtain one such word for each value of $b$. Similarly, partial flows versus $b$ yield words describing their behaviour versus $b$, one per value of $n$. For instance, in the two examples shown in Fig. \ref{fig:2}, the obtained word for $n=1$ are $-+$ (panel a) and $+$ (panel b), and $-0$ in both panels for $n=5$.

After this procedure, we obtain two indexed sets of words: one set describing behaviors versus $n$ and indexed by $b$, and one set describing behaviours versus $b$ and indexed by $n$. We concatenate all words in each set (in order of increasing $b$ or $n$) into a sentence. As in the case of words, we moreover remove consecutive repetitions of words within the sentence. 
Overall, the sentence versus $n$ is obtained by 
(1) ordering words versus $n$ by increasing value of $b$
(2) removing consecutive repetitions
(3) adding a separator $|$ between words to obtain a string.  For instance, if successive words versus $n$ are $-+$, $-+$, $-+$, $-0$, $0+$ for the values of $b$ in increasing order, the final sentence versus $n$ is 
$-\!+\!|\!-0|0+$.
The sentence versus $b$ is obtained in the same fashion. 

\begin{figure}
\subfigure[summary of the labelling procedure]{\includegraphics[width=\columnwidth]{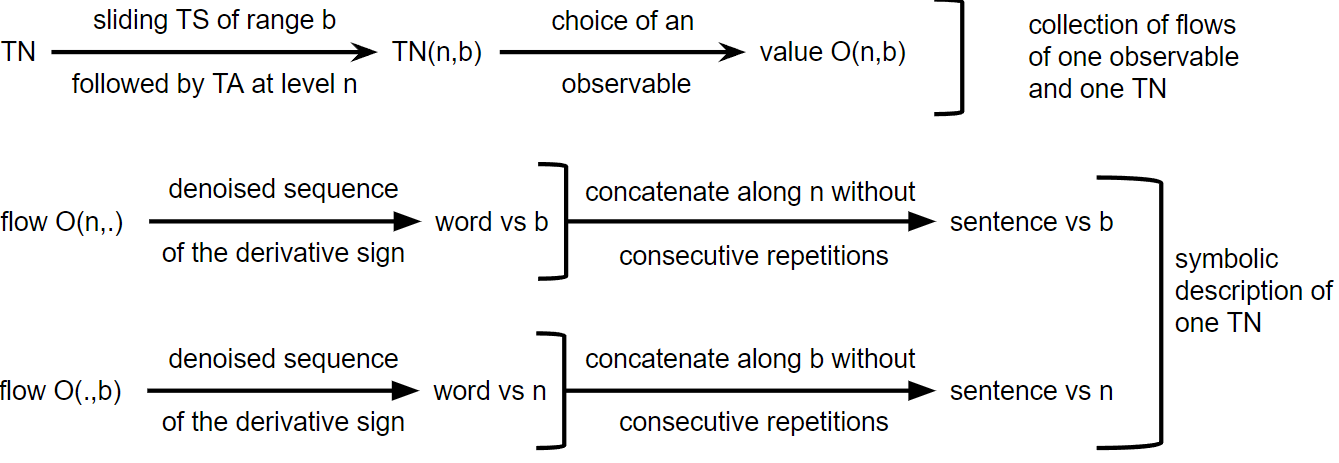}}
\subfigure[application to a particular case]{\includegraphics[width=\columnwidth]{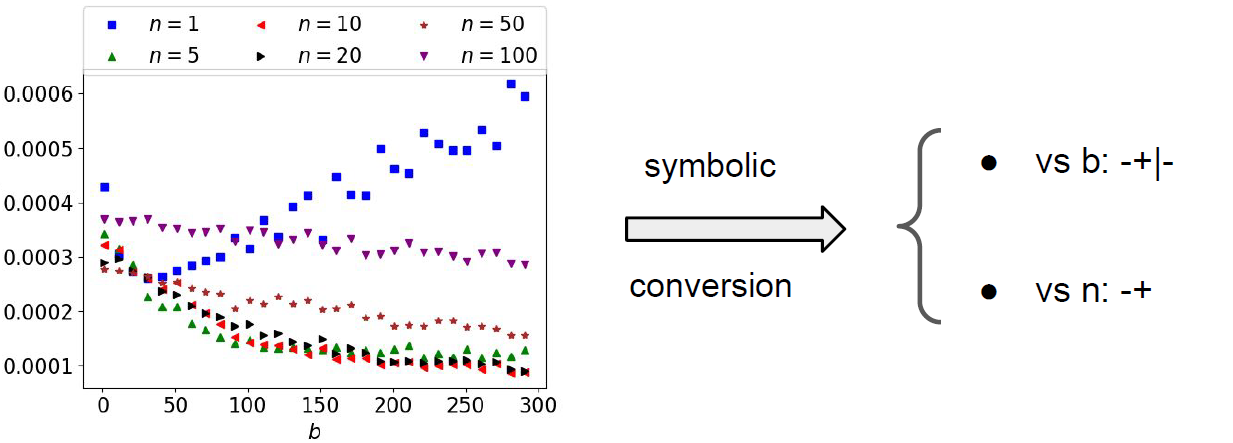}}
 \caption{\label{fig:5}\textbf{Workflow of the labelling by sentences of one TN viewed through the lens of one observable.}
    (a) Labelling workflow. Flows are averaged over ten realizations of the STS and words are extracted by a neural network.
    (b) Illustration with the ``conf16'' TN and observable ``3-NCTN\_error'' (same as in Fig.~\ref{fig:2}a). 
    To obtain the sentence versus $b$, we consider the flows versus $b$ at each value of $n$: for $n=1$ the flow is first decreasing then increasing, which yields the word $-+$;
    for larger values of $n$ the flows are always decreasing, yielding words $-$.
    As we build sentences by removing consecutive repetitions of words, we obtain $-\!+\!|-$ as sentence versus $b$ (words are separated by the symbol ``$|$'').
    The sentence versus $n$ is simply $-+$ because the flows versus $n$ are decreasing then increasing for all values of $b$.
    }
\end{figure}

We finally define the label of the temporal network under study as the couple of sentences obtained
 (one versus $n$ and one versus $b$).
Note that this label is relative to the observable whose flows are computed: we obtain one label per observable.
A summary and illustration of the labelling procedure described above are given in Fig. \ref{fig:5}.

\section{Comparing temporal networks}

The labelling procedure opens the way to various paths of investigation, as one can use them to define a similarity measure between temporal networks.
For instance one can then ask what is the performance of a given generative model by comparing the labels of its instances to the ones of empirical networks. One can also use the labels of several networks to cluster them, or to investigate the heterogeneity of a set of temporal networks. In the following we propose a concrete way to do this.

\subsection{Representation of the space of temporal networks}

We now view each instance of a temporal network through the lens of a set of scalar observables, each giving rise to a mapping from the temporal network to a label. For each observable, we can then collect all obtained labels, and build a diagram as follows. Each point in the diagram corresponds to a label, and can also be seen as the set of temporal networks sharing that label: a diagram is thus dependent on the observable considered. 
Note that such diagrams are potentially suitable to study both temporal networks and observables. Here we will focus on their use for temporal networks. 

\begin{figure}
    \centering
    \includegraphics[width=\columnwidth]{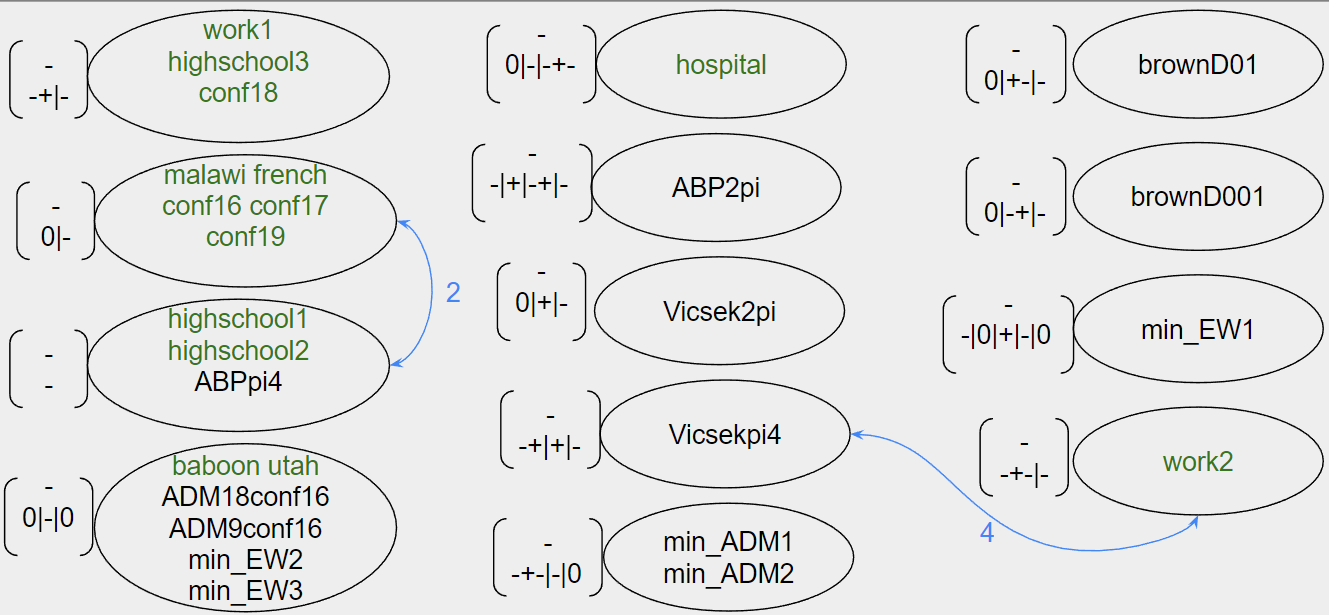}
\caption{\label{fig:6}\textbf{Diagram obtained when labelling temporal networks with respect to a specific observable.}
(namely  ``2-ECTN0sim\_trunc0vs\_n'', see Appendix \ref{sec:app_nomenclature} for the explanation of the nomenclature of observables.)
Each node of the diagram is both a label and the group of temporal networks sharing this label.
The first (resp. second) component of the label is the sentence versus $n$ (resp. $b$).
Note that for this specific observable, the same behaviour is observed versus $n$ for all temporal networks. The temporal networks considered are 
empirical data on face-to-face interactions (data set names written in green) and synthetic data generated from several models (names of models written in black). For readability, only two edges and the corresponding values of $g_{i,j}$ are shown in blue.
}
\end{figure}

Since labels are couples of sentences, a diagram can be viewed as a metric space:
the difference between two labels can indeed be simply defined as the shortest sequence of operations for rewriting one couple into the other.
If we take insertion, deletion and substitution as elementary operations, the length of this sequence is the edit distance between the two couples:
the distance between two couples of sentences $(s,t)$ and $(s',t')$ is the sum of the pairwise edit distances of their members: $d[(s,t),(s',t')]=d(s,s')+d(t,t')$.
This metric information can be made explicit by representing a diagram as a fully connected weighted directed graph, whose nodes are labelled sets of temporal networks and the edge from $i$ to $j$ bears as weight $g_{i,j}$ the length of the
shortest sequence of operations that returns $j$ if applied to $i$ (see Fig. \ref{fig:6} for an example).


\subsection{Clustering temporal networks using diagrams}

A diagram tells us how the space of temporal networks looks like when seen through the lens of a scalar observable. Several clustering are then possible, depending on whether we want compare temporal networks or classes of temporal networks.
Let us introduce useful notations to this aim:
\begin{itemize}
    \item we denote by $D(\mathcal{O})$ the diagram associated to the observable $\mathcal{O}$;
    \item we call a set or a family of temporal networks a class $C$;
    \item for $i\in D(\mathcal{O})$ a node of the diagram, we denote by $n_{C}[\mathcal{O}](i)$  the number of temporal networks from the class $C$ that belong to $i$, i.e., that are labelled by the label of the node $i$. $n_{C}[\mathcal{O}]$ is called the repartition function of the class $C$ in the diagram $D(\mathcal{O})$;
    \item $I_{\mathcal{O}}$ maps each TN to the node in $D(\mathcal{O})$ that contains it;
    \item $S$ denotes the space of observables considered; it is a finite subset of the space of all possible observables.
\end{itemize}

Each diagram provides a clustering of temporal networks (simply by grouping together the networks with the same label). However, such clusterings are observable-dependent, while we would 
like to combine information from all the considered observables in a systematic way. 
A first, naive solution would consist in considering two temporal networks to be equivalent or part of the same cluster if and only if their labels match for every observable. Such a condition is however too restrictive. 
Another possibility is to consider the space of observables as a sampling space, and a measure of similarity or distance between two networks as a random sampling experiment: specifically, to compare two networks, we draw at random an observable and we compare the labels of the two networks for the chosen observable. A difficulty of this approach is to define and justify a probability distribution over  observables. Here, for simplicity we consider a uniform distribution over all observables considered
(see subsection \ref{subsec:2} for a presentation of the observables we have used).

\subsubsection{\label{subsubsec:1}Comparing two temporal networks}

Let us consider two temporal networks $G$ and $G'$.
If we draw an observable at random, the probability that we obtain the same label for $G$ and $G'$ can be written as:
$$\frac{1}{|S|}\sum_{\mathcal{O}\in S}\delta_{I_{\mathcal{O}}(t),I_{\mathcal{O}}(t')} \ .$$
It can be interpreted as a degree of similarity between the two temporal networks:
if it equals $1$, $G$ and $G'$ always share the same labels so they behave in similar ways with respect to the reshuffling and aggregation procedures, for all considered observables. If instead this probability is equal to $0$, $G$ and $G'$ behave differently for all observables.
We call this similarity measure the \textit{raw similarity}, because it does not consider any metric structure associated with the labels.

By taking into account the metric structure, we can define another similarity measure.
First, note that we can introduce a distance between temporal networks relative to any chosen observable:
$$d_{\text{TN}}(G,G'|\mathcal{O})=g(I_{\mathcal{O}}(G),I_{\mathcal{O}}(G')) \ ,$$
where $g$ is the edit distance introduced above.
We then rescale this distance between $0$ and $1$ for each observable through:
$$\hat{d}_{\text{TN}}(G,G'|\mathcal{O})=\frac{1}{M}d_{\text{TN}}(G,G'|\mathcal{O})$$
where:
$$M=\max_{G,G'}\left(
d_{\text{TN}}(G,G'|\mathcal{O})
\right)$$
is the maximal distance encountered between two temporal networks among the set we analyze.
Assuming we restrict the analysis to a finite set of networks, $M$ is well-defined.
The rescaled distance can then be transformed into a similarity depending on the chosen observable, and we obtain a global similarity measure by averaging similarities obtained for all considered observables.
In practice, we found a better contrast by taking the geometric average than with the arithmetic one
\footnote{Note also that one could potentially define a non uniform average putting more weight on specific observables of interest.}:
$$\text{Sim}(t,t')=\left(\prod_{\mathcal{O}\in S}(1-\hat{d}_{\text{TN}}(t,t'|\mathcal{O}))\right)^{\frac{1}{|S|}}$$
We call this similarity measure the \textit{metric similarity} in the following.

Both the raw and metric similarities can be used to evaluate how far a generative model is from the empirical network it should represent. They can also be used as objective functions to be maximized by e.g. a genetic algorithm encoding a model to tune its parameter and obtain synthetic data as close as possible to given empirical data (as done e.g. in \cite{le2023modeling} with a different objective function).

Moreover, each similarity measure between temporal networks gives rise to a weighted undirected graph, where nodes are temporal networks and edge weights are similarities between nodes. Such a graph representation can then be leveraged to detect clusters of temporal networks by running any community detection algorithm on it (see subsection \ref{subsec:0} for examples using empirical and synthetic data).

\subsubsection{\label{subsubsec:2}Characterizing a class of temporal networks}

Given a class $C$ of temporal networks, such as empirical data collected in similar environments (e.g., schools), or as a set of instances of a model with different parameter values, another relevant question concerns its characterization as a class.
A first characterization can be given by the set of pairwise similarities between the temporal networks of the class.
It is however possible to go further by using the repartition functions of $C$ for different observables. Namely,
given an observable $\mathcal{O}$ and $n_{C}[\mathcal{O}]$ the associated repartition function, we define the probability for $C$ to occupy the node $i$ as
$$P_{C}[\mathcal{O}](i)=\frac{1}{|C|}n_{C}[\mathcal{O}](i)$$
and call this probability the occupancy probability of the class $C$. For each observable $\mathcal{O}$, we can then extract several properties of $C$ (we omit the label $\mathcal{O}$ to make definitions more readable):
\begin{itemize}
    \item the \textit{area} of $C$ is the number of nodes with a non-zero probability to be occupied by $C$. It is an integer bounded between 1 and $|C|$. If it equals 1 (resp. $|C|$), $\mathcal{O}$ is said to be universal (resp. specific) with respect to $C$.
    \item the \textit{diameter} of $C$ is the average distance between labels encountered in $C$: $\sum_{i,j}d(i,j)P_{C}(i)P_{C}(j)$. It indicates how different on average are two temporal networks extracted at random in $C$.
    \item the \textit{heterogeneity} is given by the entropy of $P_{C}$, which indicates how uniformly the temporal networks of $C$ are distributed among the nodes occupied by $C$.
\end{itemize}

Linking $P_{C}$ to our previous pairwise distances framework allows us to introduce a further property, which quantifies the relationship between a change in the occupancy probability and a change in labels:
$$\forall i,j\in D(\mathcal{O}),~K_{i,j}^{C}=\frac{\left|P_{C}(i)-P_{C}(j)\right|}{\left|g_{i,j}\right|}$$
The definition of $K^{C}$ is reminiscent of the definition of a derivative. Thus, small values in $K^{C}$ indicate that the occupancy probability varies smoothly with respect to the labels:
close labels contain close amounts of temporal networks belonging to $C$. From a statistical point of view, it would mean that variations in $P_{C}$ are bounded by variations in labels, indicating a correlation between the two.

\subsubsection{\label{subsubsec:5}Comparing two classes of temporal networks}

Let us consider a model of temporal networks with its parameters, and a class of empirical temporal networks, such as e.g. temporal networks describing social interactions in various conferences. An important question about the relevance of the model concerns the probability that the model produces temporal network instances similar to the empirical ones. We can quantify this by asking, when we vary the model parameters, how many of the resulting artificial temporal networks share a label with an empirical one? The corresponding overlap between the synthetic class $C$ and the empirical class $C'$ can be defined as the co-occupancy probability of the two classes:
$$\text{Overlap}(C,C'|\mathcal{O})=P(C\cap C'|\mathcal{O})=\sum_{i\in D(\mathcal{O})}P_{C}(i)P_{C'}(i) .$$
The arithmetic average of this overlap over all considered observables yields a similarity measure between classes that we call \textit{overlap similarity}
(as in previous cases, one could also define a weighted average giving more importance to specific observables):
$$\text{Overlap}(C,C')=\frac{1}{|S|}\sum_{\mathcal{O}\in S}\text{Overlap}(C,C'|\mathcal{O}) .$$
We also note that this similarity measure can be used to map classes of temporal networks into a weighted undirected network, where nodes are classes and edge weights are global overlaps between classes.

\section{Results}

We now illustrate the procedure and concepts developed above in
concrete cases, using both empirical and synthetic data sets.

\subsection{\label{subsec:2}Temporal network data sets}

We consider 27 data sets describing temporal networks, corresponding to (i) 14 publicly available empirical data sets on social interactions with high temporal resolution
\cite{barrat2013temporal,sociopatterns,toth2015role} and (ii) 13
to models of temporal networks. We consider 6 models representing the dynamics of pedestrians and their physical proximity, 4 temporal network models from the Activity Driven with Memory (ADM) class \cite{perra2012activity,Laurent_2015,le2023modeling}, and 3 ad hoc models of temporal edge dynamics.

In this sub-section, we give a brief description of the data sets:
for the empirical data, we indicate when and where the data were collected and for the models we give the principle behind them.
For more details about the sizes of the different data sets, see Appendix \ref{sec:AppendixCC}.
For more details about the models, see Appendices \ref{sec:Appendix} and \ref{sec:AppendixBB}.

Note that, while here for illustration purposes we focus on temporal networks of face-to-face interaction, our framework is applicable to any type of temporal networks, including networks of higher-order interactions \cite{battiston2020networks}.

\subsubsection{Empirical data sets}

The empirical temporal networks we use represent face-to-face interaction data collected in various contexts using wearable sensors that exchange low-power radio signals
\cite{sociopatterns,toth2015role}. 
This allows to detect face-to-face close 
proximity with here a temporal resolution of about 20 seconds \cite{cattuto2010dynamics}. All the data we used have been made publicly available by the research collaborations who collected the data \cite{sociopatterns,toth2015role}. They correspond to 
data collected among human individuals in conferences, schools, a hospital, workplaces, and also within a group of baboons.
In all cases, 
individuals are represented as nodes, and an edge is drawn between two nodes each time the associated individuals are interacting with each other. 

The data sets we consider are:
\begin{itemize}
    \item ``conf16'', ``conf17'', ``conf18'', ``conf19'': these data sets were collected in scientific conferences, respectively 
    the 3rd GESIS Computational Social Science Winter Symposium (November 30 and December 1, 2016), 
    the International Conference on Computational Social Science (July 10 to 13, 2017),
    the Eurosymposium on Computational Social Science (December 5 to 7, 2018), and the 41st European Conference on Information Retrieval (April 14 to 18, 2019) \cite{genois2022combining};
    
   \item the ``utah'' data set describes the proximity interactions which occurred on November 28 and 29, 2012 in an urban public middle school in Utah (USA) \cite{toth2015role};
    \item the ``french'' data set contains the temporal network of contacts between the children and teachers that occurred in a french primary school on Thursday, October 1st and Friday, October 2nd 2009. It is described in \cite{Gemmetto2014,10.1371/journal.pone.0023176};   
    \item the ``highschool1'', ``highschool2'' and 
    ``highschool3'' data set describe the interactions between students in a high school in Marseille, France 
\cite{10.1371/journal.pone.0107878,mastrandrea2015contact}. They were respectively collected for
 three classes during four days in Dec. 2011, 
 five classes during seven days in Nov. 2012 
 and nine classes during 5 days in December 2013;

    \item the ``hospital'' data set contains the temporal network of contacts between patients and health-care workers (HCWs) and among HCWs in a hospital ward in Lyon, France, from December 6 to 10, 2010. The study included 46 HCWs and 29 patients \cite{10.1371/journal.pone.0073970};

    \item the ``malawi'' data set contains the list of contacts measured between members of 5 households of rural Kenya between April 24  and May 12, 2012 \cite{kiti2016quantifying};

    \item the ``baboon'' data set contains observational and wearable sensors data collected in a group of 20 Guinea baboons living in an enclosure of a Primate Center in France, between June 13 and July 10 2019 \cite{gelardi2020measuring};
    
    \item the ``work1'' and `work2''
    data sets contain the temporal network of contacts between individuals measured in an office building in France, respectively from June 24 to July 3, 2013 \cite{NWS:9950811} and 
     during two weeks in 2015 \cite{Genois2018}.
\end{itemize}

\subsubsection{Pedestrian models}

Pedestrian models consist in stochastic agent-based models implemented in discrete time.
These agents move through a two-dimensional space and are point oriented particles.
In the simulations considered here, the 2D space is a square with reflecting boundary conditions.
A temporal network is built from the agents' trajectories according to a rule similar to the one used in empirical face-to-face interactions:
an interaction between two agents $i$ and $j$ is recorded at time $t$ if $i$ and $j$ are close enough and oriented towards each other at $t$.

In all pedestrian models considered in this paper, agents are point particles.
Three types of models are considered:
\begin{itemize}
    \item Brownian particles without any interaction:
    ``brownD01'' and ``brownD001''.
    These models differ only by the value of the diffusion coefficient of the agents.
    \item active Brownian particles \cite{solon2015active}:
    ``ABP2pi'' and ``ABPpi4''.
    The orientation vector of an active Brownian particle follows a Brownian motion, whereas its position vector enjoys an overdamped Langevin equation with a self-propelling force.
    This force has constant magnitude and is parallel to the orientation vector.
    In our case, however, the noise contributing to the velocity vector is zero, meaning the velocity is equal to the self-propelling force.
    Besides, the noise giving the angular velocity is a uniform random variable in $[-\theta,\theta]$.
    In ``ABP2pi'', $\theta=\pi$ and in ``ABPpi4'', $\theta=\frac{\pi}{8}$.
    \item the Vicsek model \cite{vicsek1995novel}:
    ``Vicsek2pi'' and ``Vicsekpi4''.
    In this model, the velocity of a particle at the next time step $t+1$ points in the same direction as the average velocity of its neighbours at time $t$, and an angular noise is added.
    The velocity modulus is constant and identical for every particle.
    In ``Vicsek2pi'', the velocity direction is drawn uniformly in an interval of size $2\pi$ around the average velocity of the neighbours, i.e., is completely random. In ``Vicsekpi4'', the velocity direction is drawn uniformly in an interval of size $\frac{\pi}{4}$ around the average velocity of the neighbours.
\end{itemize}

We report in Appendix \ref{sec:AppendixCC} the sizes and durations used in each model.

\subsubsection{\label{subsubsec:4}Activity Driven with Memory models}
The class of 
Activity Driven with Memory (ADM) models is an extended framework \cite{le2023modeling} of the original Activity Driven (AD) model \cite{perra2012activity}. In practice, 
an ADM model is a stochastic agent-based model in discrete time that produces a synthetic temporal network of interactions between agents.

We refer to \cite{le2023modeling} and Appendix \ref{sec:Appendix} for a detailed description of the models and their phenomenology, and provide here a brief reminder of their definition. 
In a nutshell, we consider $N$ agents, each endowed with an intrinsic activity parameter, and who interact with each other at each discrete time step in a way depending on their activity and on 
the memory of past interactions between agents.
This memory is encoded in another temporal network between the same agents, called the social bond graph: in this weighted and directed temporal graph, the weight of an edge represents the social affinity of an agent towards an other agent.
At each time step, agents thus choose partners to interact with depending on their social affinity towards other agents. The affinity is then updated by the chosen interactions through a reinforcement process: social bonds between interacting agents strengthen while the social affinity weakens if two agents do not interact.

The ADM models we considered in this paper are ``ADM9conf16'', ``ADM18conf16'', ``min\_ADM1'' and ``min\_ADM2''. Here
the numbers ``9'' or ``18'' stand for the specific dynamical rule of the model (in \cite{le2023modeling} a large number of possible variations of ADM rules has been explored) and the suffix ``conf16'' means that the parameters of the model have been tuned in order to resemble the empirical data set ``conf16''.
For more detail about those models and how their parameters have been tuned, we refer to Appendix \ref{sec:Appendix}.
We also report in Appendix \ref{sec:AppendixCC} the sizes and durations used for each model.

\subsubsection{Edge-weight models}

An edge-weight (EW) model is also a stochastic agent-based model in discrete time producing a temporal network. However, contrarily to the ADM or pedestrian models, here agents are not nodes but edges of this temporal network.

In these models, edges are independent of each other.
Their probability of activation is given by the fraction of time they have been active since their last ``birth'', which is defined either as the starting time of the temporal network, i.e. the time step 0, or as the last time the edge's history has been reset. We refer to Appendix \ref{sec:AppendixBB} for more details on the three variants we consider here, denoted  ``min\_EW1'', ``min\_EW2'' and ``min\_EW3''. See Appendix \ref{sec:AppendixCC} for the sizes used in the simulations of the models.

\subsection{\label{subsec:3}Observables}

For each data set considered, we have computed the flow of 43 observables under time aggregation and shuffling and obtained the corresponding labels.
Observables of interest in a temporal network split in \textit{scalar} observables, which yield a single realization per data set (e.g., the average degree, or the degree assortativity of the aggregated network), and \textit{distribution} observables, which can be sampled into a one dimensional probability distribution per data set (e.g., the contact durations).

\subsubsection{\label{subsubsec:distr_obs}Distribution observables}

We consider as objects both nodes or edges and, for each, their properties listed below:
\begin{itemize}
    \item \textit{duration}:
    number of consecutive snapshots the object is active, i.e. present in the temporal network;
    \item \textit{interduration}:
    number of consecutive snapshots the object is inactive, i.e. absent from the temporal network;
    \item \textit{event\_duration} \cite{Karsai2012}:
    equivalent to the duration but applied to trains of the object, it is the number of consecutive packets separated by less than two timestamps (a packet is a maximal time interval over which the object is active).
    \item \textit{time\_weight}:
    total number of snapshots the object has been active over the temporal network duration.
\end{itemize}
For each property $x$, we measure its distribution over the temporal network considered.
Since, as previously mentioned, our methodology can only handle scalar observables, we then reduce each distribution to its first modified moments $\left<x\right>$ and $\frac{\left<x^{2}\right>}{\left<x\right>}$ \cite{berlingerio2013network}  (higher moments could obviously be added to the list at will).

\subsubsection{\label{subsubsec:3}Motif-based observables}

We also consider a set of observables based on spatio-temporal motifs called Egocentric Temporal Neighbourhood motifs (ETN, see \cite{Longa2022}). Indeed, they have recently been shown to be useful tools to characterize temporal networks, and also 
to form building blocks able to decompose and reconstruct instances of temporal networks \cite{longa2022neighbourhood}.
Motifs consist in sub-temporal graphs extracted on a few consecutive timestamps.
This number of timestamps is called their depth, which is either 2 or 3 in our case.
We consider on the one hand the ETNs defined in \cite{Longa2022}, that we call NCTN, for Node-Centered Temporal Neighbourhood.
In a NCTN, only the interactions between a given node and its neighbours are taken into account. Interactions between two distinct neighbours are not considered, so that the NCTNs do not contain any information on triangles. 
We moreover use an extension of ETN, the Edge-Centered Temporal Neighbourhood (ECTN). In an ECTN, the interactions between the nodes of a given edge and the neighbouring nodes are reported: thus, ECTNs can 
include triangles to which that edge belongs \cite{lebail2023from}.
For both NCTNs and ECTNs, we compute:
\begin{itemize}
    \item the total number of motifs in the data set \textit{nb\_tot}, including repetitions of the same motif;
    \item the number \textit{nb\_diff} of distinct motifs,
     i.e. repetitions are removed;
    \item the difference \textit{motif\_error} between the frequency of an observed motif and its probability predicted under the assumption of statistical independence between its parts (see Appendix \ref{sec:AppendixD}), as this is a measure of correlations in the network.
\end{itemize}

Moreover, if we group motifs in isomorphism classes, we obtain a vector with \textit{nb\_diff} components, where each component is given by the number of occurrences of the associated motif.
We call this vector a motif vector (NCTN or ECTN vector).
We then consider the following observables:
\begin{itemize}
    \item\textit{sim ``vs $n$''} (resp. \textit{``vs $b$''}):
    cosine similarity between the motif vectors of TN$(n,b)$ and TN$(1,b)$ (resp. TN$(n,1)$);
    \item\textit{sim\_trunc}:
    same as ``sim'' but we truncate each motif vector to its 20 largest components.
\end{itemize}

\subsubsection{Other observables}

In addition to the distribution and motif-based observables described above, we computed the flows of the average instantaneous clustering coefficient (ICC), i.e., 
the average of the clustering coefficient of all snapshot networks.
We note that many other observables could be added to the list we considered here, such as, e.g., the average instantaneous node degree or the size of connected components. As our aim here is to provide a proof of concept of the whole procedure, we do not extend further our list of observables at this stage.

\subsection{\label{subsec:0}Diagrams and clusters of temporal networks}
The goal of this subsection is to show that our labelling framework can be combined with usual tools of complex systems (clustering, similarity matrices, statistical analysis, etc.) to identify clusters of temporal networks, describe these clusters by intrinsic properties and evaluate their proximity with each other (i.e. detect clusters of clusters).
For this, we apply our framework to the 27 data sets and 43 observables described in the previous subsections, and focus on temporal network analysis rather than observable analysis.

\subsubsection{Pairwise analysis}

We follow here the steps described in section \ref{subsubsec:1}: we first evaluate the proximity of each pair of temporal networks using both 
the raw similarity and the metric similarity. For each similarity, 
we build a similarity matrix ---or equivalently a proximity network--- between temporal networks, on which community detection algorithms can be run.

\begin{figure}
\subfigure[Raw similarity network.]{\includegraphics[width=\columnwidth]{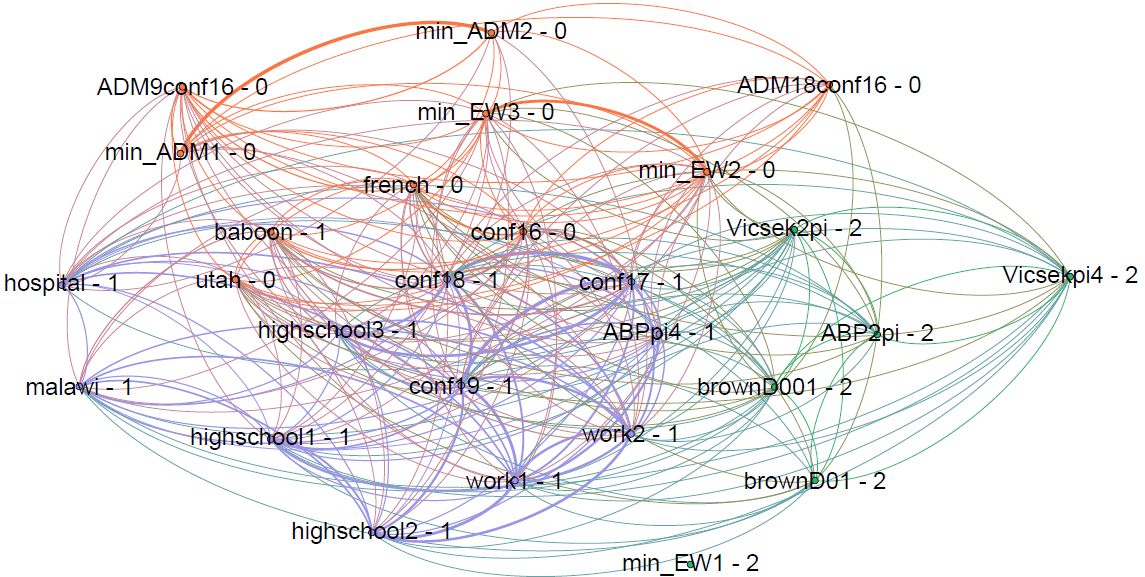}}
\subfigure[Raw similarity matrix.]{\includegraphics[width=\columnwidth]{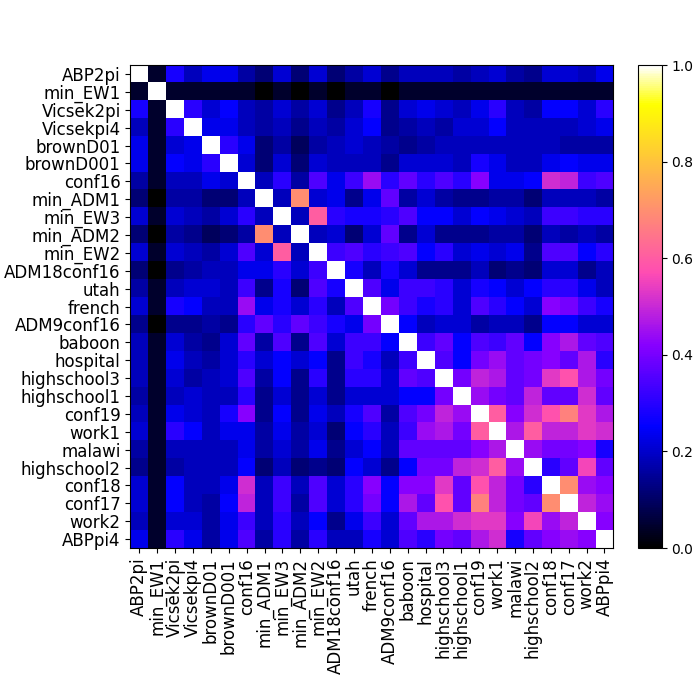}}
    \caption{\label{fig:8}\textbf{Pairwise probability to share the same label, sampled over all observables.}
    (a) We used Gephi \cite{ICWSM09154} to visualize the graph and ran a community detection algorithm based on the Louvain method.
    Edge thickness is proportional to the edge weight, which is the probability of sharing the exact same label.
    For a better readibility, only the strongest edges are visible (75.50\,\% of all edges).
    Three communities are returned by the Louvain algorithm:
    the community 0 in orange, the community 1 in purple and the community 2 in green.
    The community to which nodes belong is written at the right of their labels.
    Intra-community links are of the same colour as the community while inter-community links are plotted in gray.
    The orange community (0) contains the ADM models, a few empirical data sets and the models min\_EW2 and min\_EW3.
    The purple community (1) contains exclusively empirical data sets and the model ABPpi4.
    The green community (2) contains pedestrian models and the model min\_EW1.
    (b) Similarity matrix, from which some large similarity cases can be readily seen, such as the group of conference data, the two min\_ADM models or min\_EW2 and min\_EW3.
    }
\end{figure}

\begin{figure}
\subfigure[Metric similarity network.]{\includegraphics[width=\columnwidth]{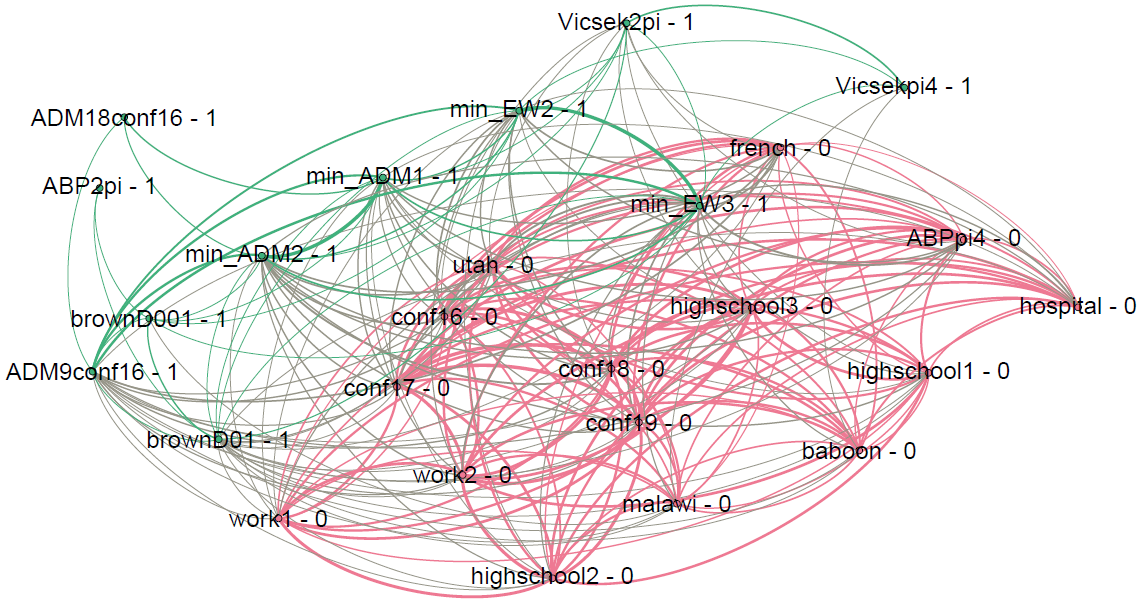}}
\subfigure[Metric similarity matrix.]{\includegraphics[width=\columnwidth]{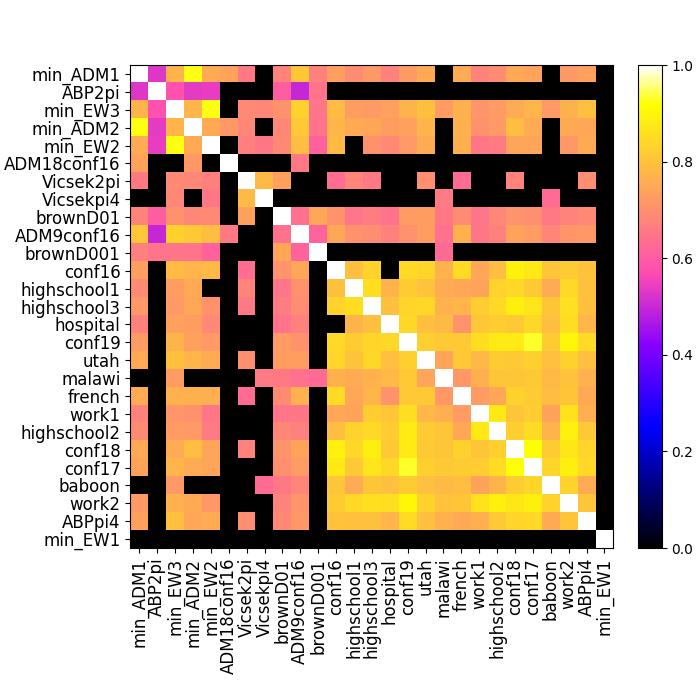}}
    \caption{\label{fig:9}\textbf{Geometric average over observables of the edit similarity between labels.}
    (a) As in Fig. \ref{fig:8}, we used Gephi to visualize the graph and ran the same community detection algorithm.
    The min\_EW1 model is missing, because it has zero similarity with every other TN.
    Two communities of comparable size are detected, respectively populated with empirical data sets (0, orange) and models (1, green).
    (b) The two detected communities are visible on the similarity matrix.
    Note that these communities do overlap, which is expected since most of the models considered here aim at reproducing some properties of the empirical data sets.
    }
\end{figure}

The resulting proximity networks and similarity matrices obtained  are shown in Fig.~\ref{fig:8} for the raw similarity case and in Fig.~\ref{fig:9} for the metric similarity. To extract groups of temporal networks, we apply the Louvain algorithm to the proximity networks.
Note that there is a resolution parameter in the Louvain algorithm, and changing it results in different communities.
We thus used the following criterion:  we measure the rate of change of the partition into communities as we change the resolution parameter by computing 
the adjusted Rand score \cite{scikit-learn} of similarity between partitions obtained at two consecutive values of increasing resolution. Then we selected the resolution maximizing the adjusted Rand score. If a plateau was observed, we chose the resolution maximizing the modularity.

From Fig.~\ref{fig:8}, we see that considering raw diagrams allows to separate between ADM and pedestrian models, while most empirical networks form a third class. The 
 ``conf16'' empirical data is classified with the ADM models, which is probably due to the fact that the ADM models have been tuned in order to resemble the ``conf16'' data set.
A different structure is obtained by taking the metric similarity, as shown in Fig.~\ref{fig:9}: we then obtain two clusters corresponding respectively to the synthetic and empirical temporal networks, with one exception as the model ABPpi4 belongs to the same cluster as empirical data sets.

We also note that the model ``ADM9conf16'', put forward in  \cite{le2023modeling}, reproduces well the empirical distributions of many observables (distributions of contact and intercontact duration, ETN motifs, etc) of the ``conf16'' data set. 
As seen in Fig.~\ref{fig:9}b, 
our approach indeed finds it closer to the empirical data sets 
than the model ``ADM18conf16'', which had a lower performance (see \cite{le2023modeling}) at reproducing the empirical properties.
However, ``ADM9conf16'' still does not belong to the same cluster as the empirical data sets: this illustrates how the mere reproduction of statistical distributions of observables of one empirical temporal network does not ensure that the label of the temporal network will also be shared. Although the labelling procedure entails only limited information (since we use only information about qualitative shapes of the flows), the fact that it contains information about all scales at once seems to 
make it more informative about a temporal network than a collection of distributions sampled at a single resolution level. Said otherwise, there seems to be more information in the way an observable changes under the aggregation and shuffling transformations, than in the precise realizations of this observable.

\subsubsection{Sensitivity analysis}

The clusters of temporal networks we obtained above depend a priori on the set of observables we choose to compute the similarity between networks. As the framework is flexible and can be extended to an arbitrary large number of observables, one can hope that the partition into clusters becomes stable as the number and diversity of observables considered becomes large enough.
To check this, we computed the evolution of the structure of communities yielded by the raw and metric similarities as the number of observables considered increased from 1 to 43.
More precisely, for each $1\leq s\leq43$,
we draw $s$ observables at random among the 43 available, 
and compute the similarity network and the resulting community structure. We then compute the adjusted Rand score $R_{s}$ between the partition in communities obtained with $s$ observables and the one obtained with $s+1$ observables.
This measure indeed quantifies the similarity between two partitions of the same set of elements (here the temporal networks), and we average its value over $100$ realizations of the random choice of observables.

\begin{figure}
\subfigure[raw similarity case]{\includegraphics[width=0.48\columnwidth]{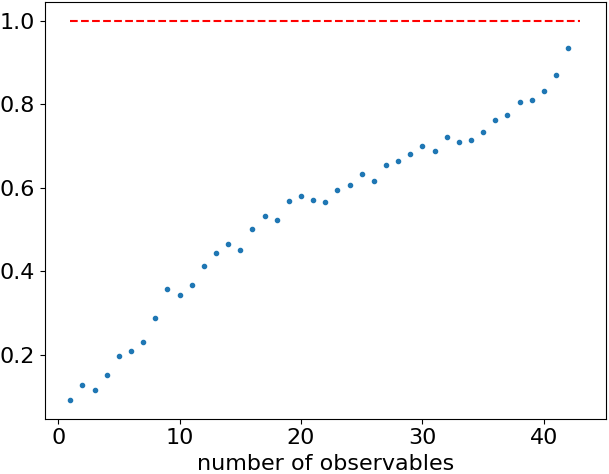}}
\subfigure[metric similarity case]{\includegraphics[width=0.48\columnwidth]{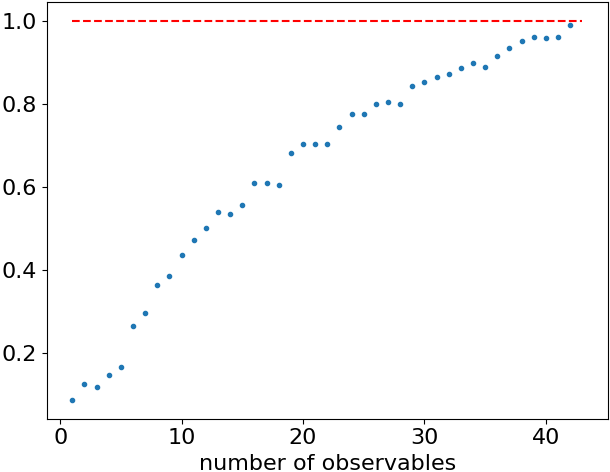}}
    \caption{\label{fig:16}\textbf{Stability of communities with respect to the number of considered observables.}
    The adjusted Rand score (displayed on the $y$-axis) increases with the number of considered observables ($x$-axis), meaning that the detected communities become increasingly stable as more observables are used. Values close to $1$ are obtained for both the raw (a) and metric (b) cases.
    }
\end{figure}

Figure \ref{fig:16} shows that $R_{s}$ increases with $s$, i.e., the structure in communities becomes more stable as more observables are added. A larger stability is obtained with the metric similarity than with the raw one, and very large values
of the Rand index are obtained when more than 30 observables are considered.
To reach a perfect stability, some additional observables might need to be considered, but the results of Fig.~\ref{fig:16} indicates that the clusters identified in Figs.~\ref{fig:8} and \ref{fig:9} are already very stable with respect to the choice of observables.

\subsubsection{Analysis of classes}

\begin{figure*}
\subfigure[Diameter histograms.]{
\includegraphics[width=0.48\columnwidth]{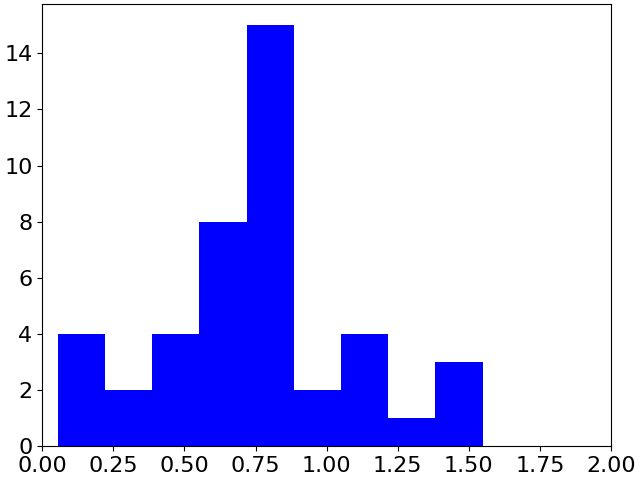}
\includegraphics[width=0.48\columnwidth]{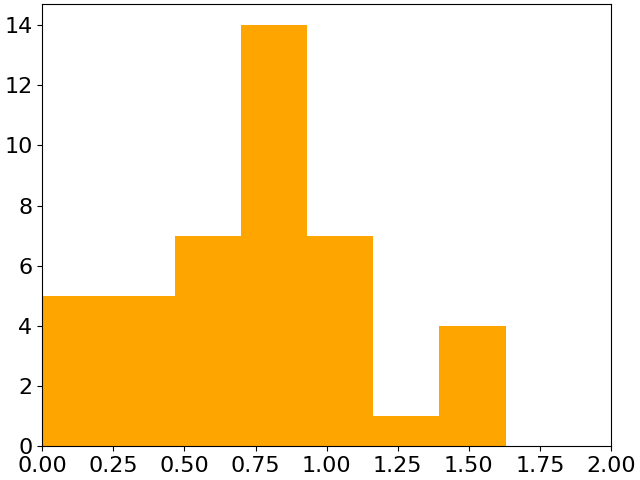}
    \includegraphics[width=0.48\columnwidth]{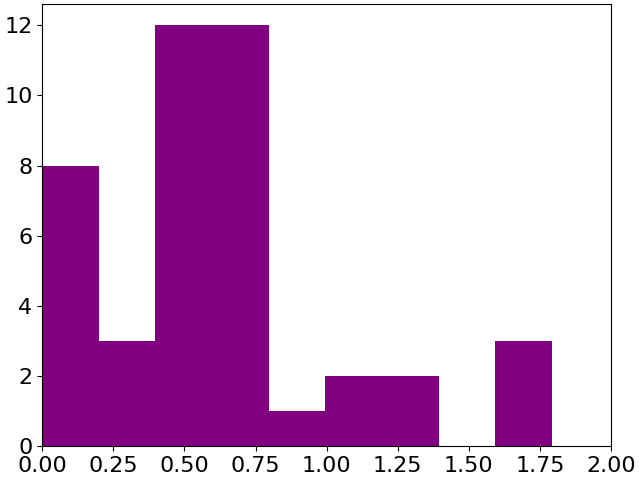}
    \includegraphics[width=0.48\columnwidth]{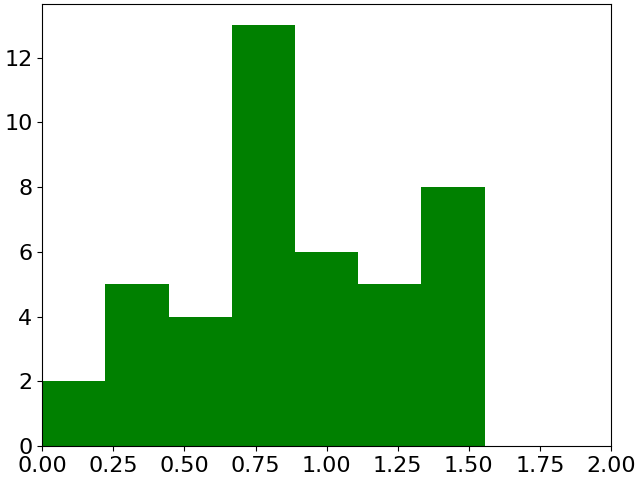}}
\subfigure[Heterogeneity histograms.]{
    \includegraphics[width=0.48\columnwidth]{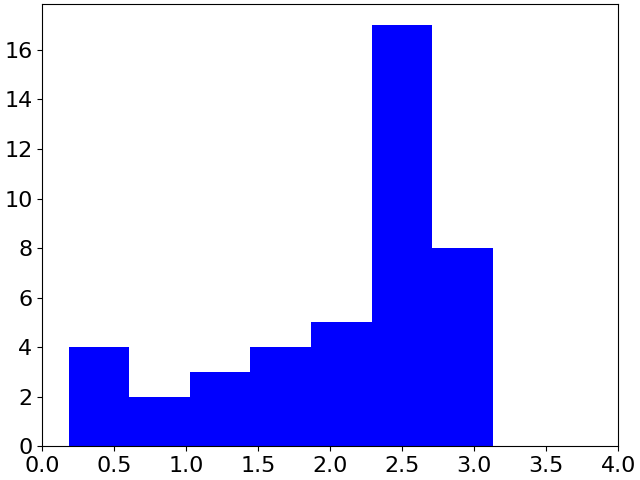}
    \includegraphics[width=0.48\columnwidth]{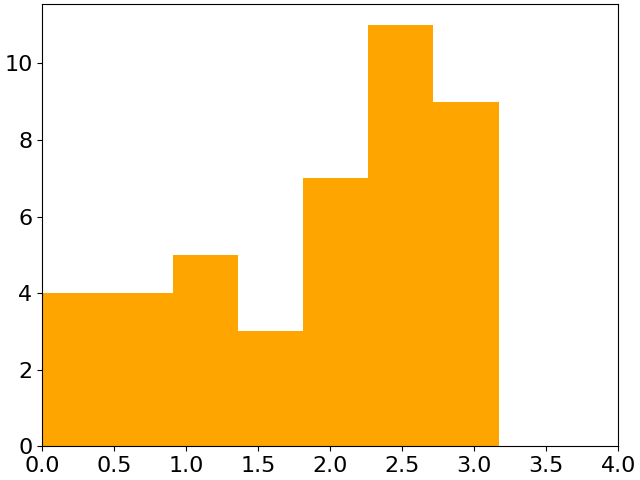}
    \includegraphics[width=0.48\columnwidth]{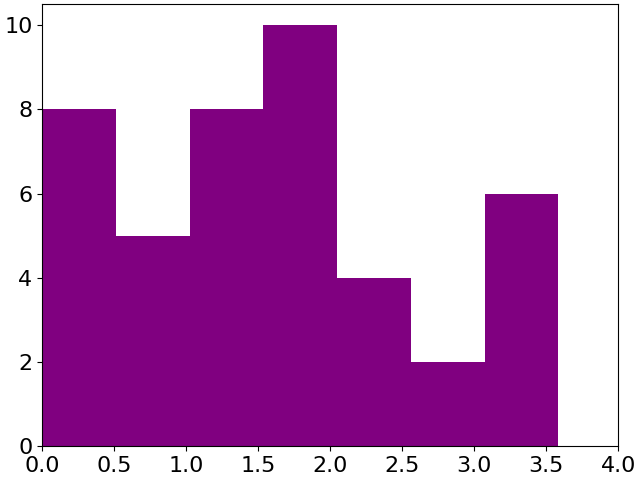}
    \includegraphics[width=0.48\columnwidth]{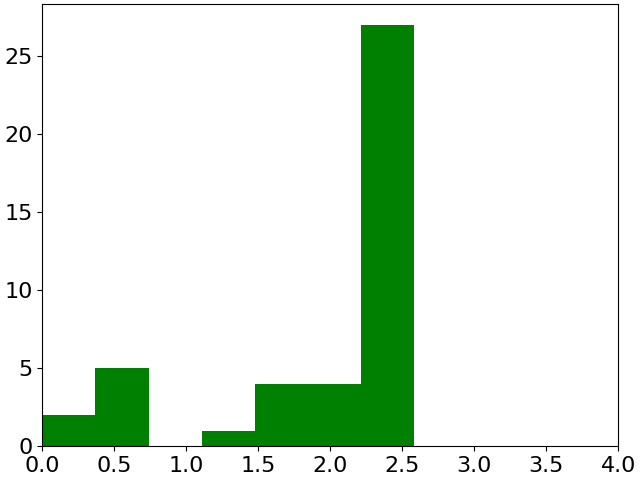}}
    \caption{\label{fig:10}\textbf{Histograms of heterogeneity and diameter.}
    For each class, the heterogeneity (a) and diameter (b) are computed from the repartition function and definitions given in sub-subsection \ref{subsubsec:2}.
    The $y$ axis indicates the number of occurrences of the corresponding value reported on the $x$ axis.
    The colors indicate the considered class following the coloring of Fig.~\ref{fig:8}a:
    orange for the class 0 (mostly ADM models), purple for the class 1 (mostly empirical data) and green for the class 2 (mostly pedestrian models).
    Histograms in blue are obtained from the average over 200 independent choices of a group of 6 temporal networks at random among the 27 available.
    }
\end{figure*}

We consider three classes of temporal networks, corresponding to the three detected communities on figure \ref{fig:8}:
\begin{itemize}
    \item class 0 (orange, 9 elements):
    ``min\_EW2'', ``min\_EW3'', ``ADM9conf16'', ``ADM18conf16'', ``min\_ADM1'', ``min\_ADM2'', ``conf16'', ``french'', ``utah'';
    \item class 1 (purple, 12 elements):
    ``ABPpi4'', ``highschool1'', ``highschool2'', ``highschool3'', ``conf17'', ``conf18'', ``conf19'', ``work1'', ``work2'', ``malawi'', ``baboon'', ``hospital'';
    \item class 2 (green, 6 elements):
    ``min\_EW1'', ``Vicsekpi4', ``Vicsek2pi'', ``brownD01'', ``brownD001'', ``ABP2pi''.
\end{itemize}
We compute the diameter and heterogeneity of each class (see section \ref{subsubsec:2} for definitions) for each observable, and display in Fig.~\ref{fig:10} the histograms of these quantities sampled over observables.
On the same figure, we also display the histograms obtained for a group of 6 temporal networks taken at random among the 27 available, averaging over 200 independent realizations of this random class.

Figure \ref{fig:10} shows that each class is characterized by both different diameter and heterogeneity distributions.
The class 1, mostly composed of empirical data sets, tends to have lower diameters, indicating that labels of empirical data sets are  close to each other for most observables. 

The heterogeneity distribution indicates how probable it is to pick a specific or a universal observable with respect to the class considered when choosing an observable at random. If the class is built randomly (blue histograms), observables tend to be specific (higher values of heterogeneity more represented than small values), i.e., have different labels for different temporal networks. Class 1 (purple) shows a clearly different behaviour, with more observables having a low heterogeneity (being more universal within the class). This reflects the fact that empirical data sets tend to share many statistical properties.

On the other hand, the class of pedestrian models (class 2, green) has almost only specific observables among the ones we considered.
This is in accordance with the similarity matrix of Fig.~\ref{fig:8}b, which shows that models from this class do not share the same labels. This might be due to two reasons:
(1) The observables' flows obtained for observables of the temporal networks created by pedestrian models are mainly flat (to the human eye) but actually noisy; the automatic attribution of labels by the neural network we built might then fail to identify the flat behaviour and instead assign complex 
 labels to these flows. For example, it may assign something like ``$-|\!+\!-\!+\!|0$'' to a flow instead of ``0''.
Thus, random complex labels may be assigned to the pedestrian models, resulting in different labels for almost every temporal network. We have checked by hand that this was not the case.
(2) The pedestrian models do not form a well-defined class because their fundamental mechanisms differ from one model to the other. 
Moreover, changing parameter values can strongly affect the model's behaviour. For instance, ``Vicseckpi4'' and ``Vicseck2pi'' are in two distinct parts in the phase space of the Vicseck model \cite{vicsek1995novel}. Interestingly, our labelling procedure seems thus to be able to distinguish between realizations of a given model corresponding to different phases and thus to have the potential to detect phase transitions undergone by a model.
Note also that the situation is different for the other models we consider, such as e.g.  ``min\_EW2'' and ``min\_EW3'', or  ``min\_ADM2'' and ``min\_ADM3'': here changing the values of the models' parameters does not affect strongly the model behaviour, and we indeed obtain very close labels.

\subsubsection{Co-occupancy of classes}

The three classes of temporal networks have different diameter and heterogeneity distributions. Do these differences reflect in non-overlapping labels?
To answer this question, we compute the overlap between our three classes, as defined in section \ref{subsubsec:5}:
recall that the overlap similarity between two classes measures the probability that two members drawn at random from these classes share the same label.
We obtain the following values:
\begin{itemize}
    \item $P(C_{0}\cap C_{1})\simeq0.24$
    \item $P(C_{0}\cap C_{2})\simeq0.15$
    \item $P(C_{1}\cap C_{2})\simeq0.18$
\end{itemize}
In particular, we see that the ADM class (class 0) has a stronger overlap with the class of empirical data (class 1) than the class of pedestrian models (class 2), and the pedestrian and ADM models almost do not overlap with each other, indicating that their statistical properties fundamentally differ from each other. This qualitative difference may be due to the role of space: in pedestrian models, agents are constrained by a 2D motion while in ADM models, the social network alone is considered.

\section{Conclusion}

In this article, we have proposed a systematic procedure to associate discrete labels to temporal networks in order to provide a way to compare instances or whole classes of temporal networks. This can help not only to validate models but also to assess the heterogeneity of temporal networks within a class of models or within an ensemble of empirical data sets. 

\begin{figure*}
    \subfigure[similarity between empirical and randomized data sets]{\includegraphics[width=\columnwidth]{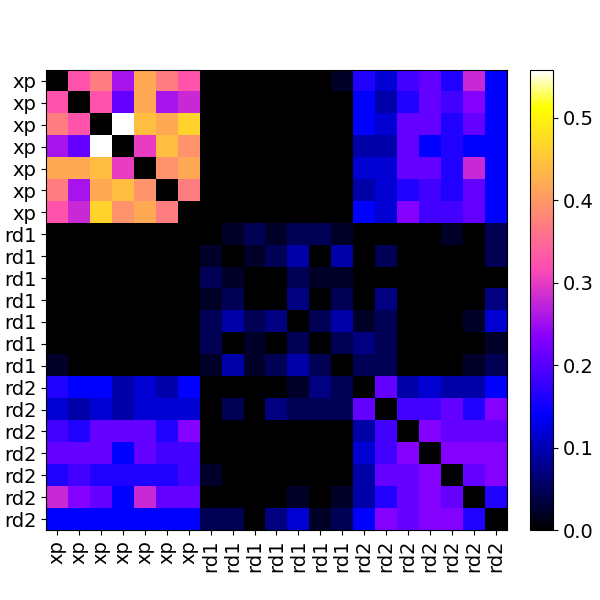}}
    \subfigure[similarity between randomized data sets]{\includegraphics[width=\columnwidth]{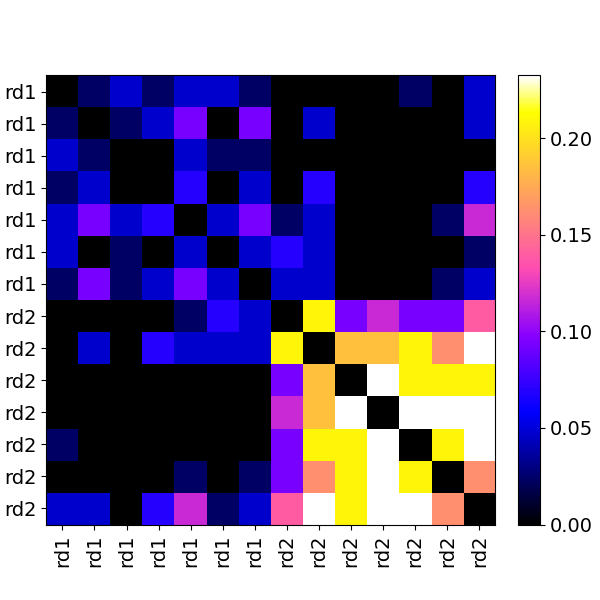}}
    \caption{\label{fig:18}\textbf{Raw similarity matrices between empirical data sets and their randomized versions}.
    For a better readability, the matrix diagonal has been filled with zeros and we considered only one empirical data set of each type:
    conference, village, groups of baboons, highschool, primary school, hospital and workplace.
    Moreover, data set have been renamed to three categories:
    (1) "xp" for any empirical data set
    (2) "rd1" (resp. "rd2") for any data set randomized according to the method 1 (resp. 2), which we describe now.
    The randomization 1 preserves only the fully aggregated network:
    it consists, for each edge, in re-drawing at random its times of activation within the network duration.
    The randomization 2 preserves both the fully aggregated network and the number of interactions at each time step.
    (a) We see that empirical data sets and their randomized versions are well separated.
    (b) The randomized versions of the empirical data sets considered are also separated according to their type, or degree (recall "rd1" is stronger than "rd2").
    Overall, we note that the stronger the randomization, the weaker the similarity between data sets.
    This is because the neural network assigning labels to flows is not perfect: it tends to denote flat but noisy flows by random lengthy labels (like e.g. '-+-' instead of '0').
    As we observe that flows are flattened by a strong randomization, it follows that their labels are random-like, making them diverse and thus different from each other.
    }
\end{figure*}

To this aim, we have considered how observables evolve when the temporal network data is transformed under a reshuffling at a certain scale, leading to a partial removal of short-time correlations in the temporal network, followed by a temporal aggregation at another scale. For simplicity, we have defined labels as describing the successive trends of the flows of observables under these transformations, without encoding their specific values. Although they thus encode only qualitative information, these labels make it possible to define a metric to compare temporal networks, and thus to evaluate a model performance, identify clusters of temporal networks and characterize these clusters by new properties (diameter, heterogeneity, etc). 
We have for instance shown how the procedure is able to separate empirical data sets from synthetic data created by models, even if some of these models have been tuned to reproduce several properties of the data. This also highlights how current models need to be improved to take into account non-trivial temporal correlations. As a further illustration of the method, we show in Figure \ref{fig:18} that the method can separate empirical data sets from their reshuffled versions, and even separate among the reshuffling methods.
To quantitatively confirm this and explore further the ability of the method to distinguish between types of reshuffling, we would need (1) to test our method with a larger set of reshuffling methods \cite{gauvin2022randomized} and (2) to improve the performance of our neural network. Indeed, the current version of the neural network currently still struggles to assign flat labels '0' to noisy flows that look flat to the human eye. 
One consequence of this limitation is that
the stronger the randomization, the weaker the similarity, as seen in Figure \ref{fig:18}.

The framework we have put forward is flexible and could be extended at will. For instance, one could consider other reshuffling procedures \cite{gauvin2022randomized} to create the flows of temporal networks and observables.
We have also here considered a certain list of observables, chosen in part arbitrarily. Depending on the context of the temporal networks considered, a different set of observables could be used.
The set of observables could also be extended, for instance including higher moments of the distributions \cite{berlingerio2013network}, or other properties of network snapshots such as average degree, assortativity or sizes of connected components. The similarity metrics could also be defined using a non-uniform average over observables if some are deemed more important than others in specific contexts.

Our work suggest some avenues for future work. 
First, the performance of the neural network we used to assign a label to a curve could probably be improved.
Second, the study could be extended to a larger and more diversified set of empirical and temporal networks, and potentially therefore also to more observables.
As briefly mentioned, the procedure can also give rise to a study focused on the observables themselves:
What are the possible or most probable labels for a given observable? Which observables yield the same or similar diagrams?
Finally, labels in themselves could be studied in more detail, revealing distinct and well-defined properties for temporal networks and observables (universality, specificity, etc.).
This would also raise many theoretical questions about the mechanisms behind the rich phenomenology of the flows: indeed,
even seemingly simple models like the min\_ADM models of this paper can exhibit large labels if an appropriate choice of their parameters is made.

In conclusion, the study of the flows of temporal networks and attached observables under partial reshuffling and aggregation is a promising tool for the study of temporal networks, which 
contains valuable information to compare, cluster and characterize temporal networks.

\section*{Acknowledgments}
This work was supported by the Agence Nationale de la Recherche (ANR) project DATAREDUX (ANR-19-CE46-0008).

\newpage
\bibliography{main}

\clearpage
\newpage

\appendix

\setcounter{figure}{0}
\setcounter{equation}{0}
\setcounter{table}{0}
\setcounter{section}{0}
\renewcommand{\thefigure}{S\arabic{figure}}
\renewcommand{\thesection}{S\arabic{section}}
\renewcommand{\thetable}{S\arabic{table}}
\renewcommand{\theequation}{S\arabic{equation}}

\widetext

\section{Disjoint and sliding time shuffling}

\subsection{\label{sec:AppendixAA}Noise removal by STS}

\begin{figure}[b]
\subfigure[disjoint time shuffling (DTS)]{
    \includegraphics[width=0.45\columnwidth]{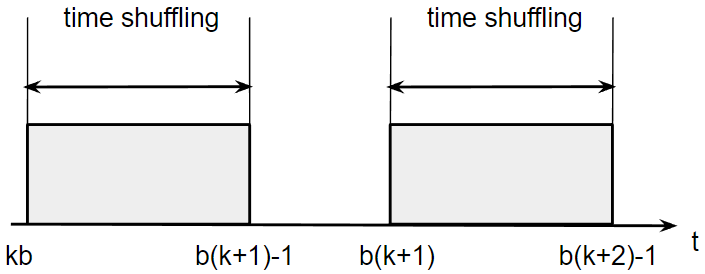}}
\subfigure[example of a noisy flow under DTS]{
    \includegraphics[width=0.45\columnwidth]{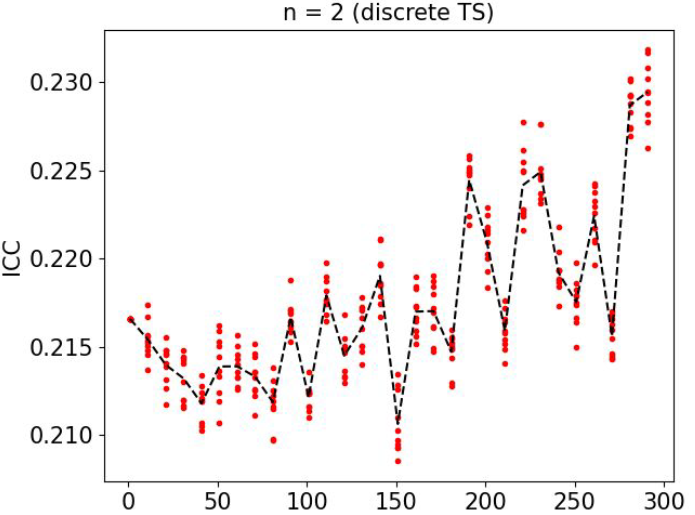}}
    \caption{\label{fig:1}\textbf{Illustration of disjoint time shuffling and a noisy flow.}
    (a) To perform a disjoint time shuffling, the timeline of the temporal network is split in successive disjoint blocs of length $b$, where $b\in\mathbf{N}$ is the parameter of the transformation. Then a time shuffling is performed within each bloc. This removes any correlation on time scales shorter than $b$ but preserves correlations on time scales larger than $b$.
    (b) Flow of the instantaneous clustering coefficient (``ICC'') under DTS for the aggregation level $n=2$ and the TN ``conf16''.
    The red dots stand for different realizations of each time shuffling while the black dashed line is the average of these realizations.
    The dispersion of the red dots appears to be smaller than the dashed line discontinuities, indicating a source of randomness that is present in the TN itself.
    }
\end{figure}

As explained in the main text, we want to compute the sign sequence of the derivative of a given curve (given in our specific case by the evolution of an observable $\mathcal{O}$ under shuffling or aggregation when the corresponding time scale $b$ or $n$ is varied).
As we are dealing with noisy data or random realizations of models, each curve will contain a certain amount of noise.
To obtain the sign sequence automatically, we need thus first to reduce the noise level as much as possible.
Therefore, we average each value $\mathcal{O}(n,b)$ over ten realizations of the local time shuffling with range $b$.
If we use a time shuffling on successive, disjoint time windows (see Fig.~\ref{fig:1}a for illustration), it can however happen that the noise of the curve is of greater amplitude than the noise originating from the time shuffling.
We show an example of such a noisy flow in Fig.~\ref{fig:1}b.

\begin{figure}
    \includegraphics[width=\columnwidth]{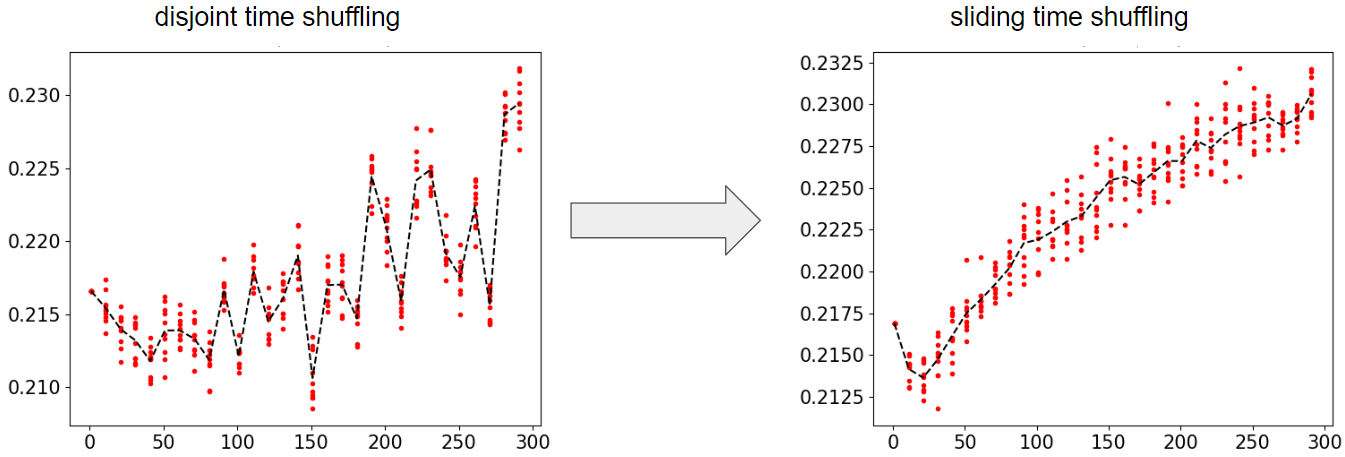}
    \caption{\label{fig:4}\textbf{Noise reduction by sliding time shuffling.}
    We consider the observable ICC at aggregation level $2$, like in Fig.~\ref{fig:1}b.
    We see that using STS instead of DTS removes most of the spurious oscillations in the ICC flow versus $b$.
    Indeed, the fluctuations of the black line are now fully explainable by the dispersion of the red dots.
    Note that the change of variation in the flow is now located at earlier $b\simeq25$, because the STS range is $2b$ instead of $b$ for DTS.
    }
\end{figure}

These oscillations are a priori surprising because the only source of randomness is precisely the local time shuffling.
It means that this additional noise is not random but might contain information. However, this information is too specific to the TN. Indeed, this pseudo-noise originates from the combination of (1) the split of the TN timeline in disjoint blocks and (2) the temporal fluctuations of the edge activity of a TN, i.e. the number of active ties at a given timestamp.
To see this, consider the toy situation:
of a temporal network with timeline for the edge activity $E(t)$ (which denotes the number of active ties at time $t \ge 0$):
 $$E(t)=E_{0}+\Delta\theta(t-t_{0})$$
($\theta$ denotes the Heaviside function).
Then if $b$ divides $t_{0}$, snapshots with different levels edge activities are not mixed by the reshuffling at scale $b$.
However, if $b$ is not such a divider, snapshots from different states do mix, which results in a different realization for most observables. This in turn causes spurious oscillations in the observables' flows.
To avoid these oscillations, we thus consider a sliding window for the shuffling procedure, as described in the main text, instead of disjoint windows. 
More precisely, the transformation we propose, that we call a sliding time shuffling (STS, Fig.~\ref{fig:0}), consists in shuffling the snapshots in the block 
$\llbracket t,t+b-1\rrbracket$ when $t$ goes iteratively from 0 to $T-b$, $T$ being the total number of snapshots.
Figure \ref{fig:4} shows that the STS indeed results in a smoothing of the observables' flows.

\subsection{\label{sec:AppendixE}Range of sliding time shuffling}

The sliding shuffling is not strictly local anymore, since there is now a non-zero probability that snapshots distant by more than $b$ get exchanged.
What is the range of the STS compared to the DTS (Disjoint Time Shuffling)? To answer this,
we need to compute the average distance travelled by a snapshot as we apply successive shufflings.

Let us define $t_{k}$ the snapshot location after $k+1$ shuffling operations. Considering a snapshot with initial position $1\leq t_{-1}\leq b$, we have $1\leq t_{0}\leq b$.
More generally:
$$1\leq t_{k}\leq b+k,\forall k\geq0$$
If $t_{k}\leq k+1$ then $t_{k'}=t_{k},\forall k'\geq k$.

Let us define $K$ as the smallest $k$ which satisfies $t_{k}\leq k+1$.
$K+1$ gives the final position of the snapshot initially located between times 1 and $b$.
What is the law of $K$? We can write
$$P(K=0)=\frac{1}{b}$$
$$P(K=1)=\frac{b-1}{b}\frac{1}{b}$$
$$P(K=k)=\left(\frac{b-1}{b}\right)^{k}\frac{1}{b}
\simeq\exp\left(-\frac{k}{b}\right)\frac{1}{b}$$

We obtain, introducing $q=\frac{b-1}{b}$:
$$\left<K\right>=\sum_{k=1}^{\infty}kP(K=k)
=\frac{1}{b}\frac{d}{dx}|_{x=1}\left(\frac{1}{1-qx}\right)$$
$$\left<K\right>=\frac{1}{b}\frac{q}{(1-q)^{2}}=bq=b-1$$

Thus, on average, a snapshot moves from the location $t_{0}$ by the same distance as when local TS is considered.
Since the initial location is $t_{-1}$ and not $t_{0}$, the STS range is $2b$ instead of $b$.
The law of $K$ being short-tailed, we can conclude that sliding time shuffling is still a local transformation, of range twice the range of local time shuffling.

\section{\label{sec:Appendix}ADM models}

The ADM models are agent-based models in which a fixed population of $N$ agents interact following specific rules at each time step.
Specifying an instance of the ADM class comes in two steps:
(1) we choose the dynamics for the agents, i.e. the rules for their behaviour;
(2) we choose values for the parameters associated to these rules.
For example, if we choose that agents can lose memory of all their past interactions at each time step (rule), we have to precise what is the probability of such an event (parameter).

Once dynamical rules for the agents have been set, two options are available to specify values for the parameters.
Instead of entering them arbitrarily, it is indeed possible to 
use a genetic algorithm to tune them automatically, so that our model produces a temporal network as similar as possible to a specified target, with respect to a set of observables. This is the procedure put forward in \cite{le2023modeling}.

\begin{figure}
\subfigure[tuning $p_{d}$ for min\_ADM1]{
    \includegraphics[width=0.4\columnwidth]
    {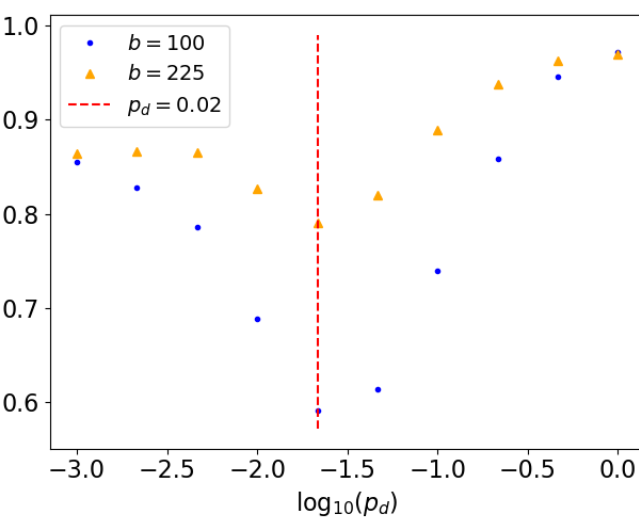}}
\subfigure[tuning $p_{d}$ for min\_ADM2]{
    \includegraphics[width=0.4\columnwidth]
    {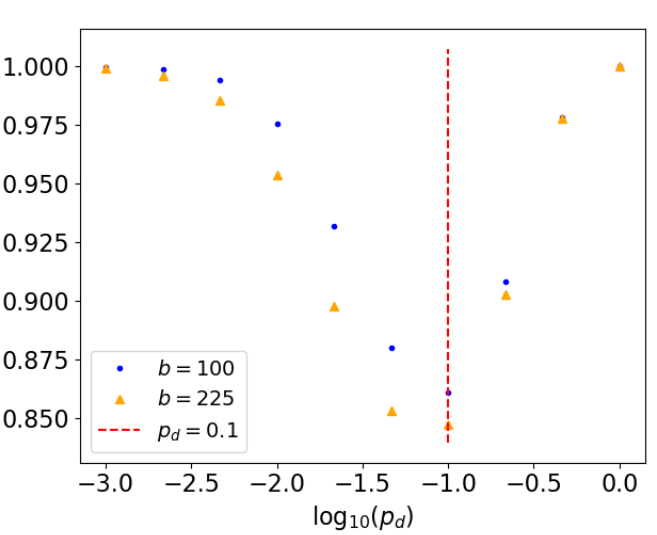}}
    \caption{\label{fig:14}\textbf{Choice of the probability $p_{d}$ of deleting memory for the models min\_ADM1 and min\_ADM2.}
    We plot the ETN similarity between TN$(n,b)$ and TN$(n,1)$ for large values of $b$ as a function of $p_{d}$.
    The aggregation level $n$ differs in the two models.
    In each case, a minimum similarity is achieved for an intermediate value of $p_{d}$, corresponding to a maximum sensitivity under STS.
    (a) min\_ADM1: $n=10$ and we obtain $p_{d}=0.02$;
    (b) min\_ADM2: $n=1$ and we obtain $p_{d}=0.1$
    }
\end{figure}

The ADM models we considered in this paper are called 
``ADM9conf16'', ``ADM18conf16'', ``min\_ADM1'' and ``min\_ADM2''.
The numbers ``9'' and ``18'' stand for two different dynamical rules (a broad variety of rules has been studied in \cite{le2023modeling}) and the suffix ``conf16'' means that the parameters of have been obtained by genetic tuning in order to resemble the empirical data set ``conf16''.
On the other hand, most parameters of ``min\_ADM1'' and ``min\_ADM2'' have been chosen by hand (given in Table \ref{tab:1}).
Only one parameter has been tuned, but without using any genetic algorithm.
This parameter is denoted by $p_{d}$ and corresponds to the probability for an agent to clear its memory of past interactions at each time step.
For both models ``min\_ADM1'' and ``min\_ADM2'', we choose the value of $p_{d}$ to maximize the sensitivity to STS at a certain, arbitrarily chosen level of aggregation.
Said otherwise, we wanted models that are not invariant under STS.
Indeed, being invariant under STS means having no time correlations, which is not an interesting situation for a temporal network. 
We thus quantified, for a given aggregation level $n$, the degree of invariance of a TN under STS as the cosine similarity between $\mathcal{O}(n,b)$ and $\mathcal{O}(n,1)$ in the limit $b\gg1$, using as observable $\mathcal{O}$ is a vector indexed by spatio-temporal motifs (its component $\mathcal{O}_{i}$ is the number of occurrences of the spatio-temporal motif $i$), namely
the Egocentric Temporal Neighbourhoods of depth 3 (ETN, see \cite{Longa2022}), written as ``3-NCTN'' in our nomenclature for observables, see Fig.~\ref{fig:14}.


Let us now describe the rules for the agents' dynamics in the four ADM models considered here.
We also refer to \cite{le2023modeling} for more details and results on these models.
Each agent $i$ is endowed with an activity parameter $a_i$ that will determine its level of activity, taken from a certain distribution, or set to the same value for all agents.
We denote by $G$ the temporal network of their interaction, called the \textit{interaction graph}.
In the four models, the $N$ agents keep a long-term memory of their past interactions, encoded in another temporal network, called the \textit{social bond graph} and denoted by $B$.
$B_{i,j}(t)$ is the directed strength of the social affinity of the agent $i$ towards the agent $j$.
At the initial time, the $N$ agents are initialized with an empty social bond graph.

Before going into more details, let us precise that the models ``ADM18conf16'', ``min\_ADM1'' and ``min\_ADM2'' share similar dynamics, while the model ``ADM9conf16'' is more complex.
Thus we will describe first the simpler models and then the model ``ADM9conf16''. 

\subsection{Models ``ADM18conf16'', ``min\_ADM1'' and ``min\_ADM2''}

At time $t$, the snapshot $G(t)$ is built as follows:
\newcounter{ADM}
\begin{enumerate}
    \item We determine which agents will be ``active'', i.e. able to propose an interaction to other agents.
Every agent has the same probability to be active, denoted by $a$.
    \item Active agents try to interact with a given number of other agents. This number is denoted by $m$ and is the same for every agent.
    \item For each proposal, the active agent will either pick a partner among its memorized contacts (probability $1-p_{g}$) or will grow its egonet, i.e. will pick an unknown partner (probability $p_{g}$).
    \item Two options:
    \begin{enumerate}
        \item To grow its egonet, two options are available to the agent: with probability $p_{u}$ it will pick a partner at random among the agents it does not know. Otherwise (so with probability $1-p_{u}$), it will grow his egonet by triadic closure.
        \item If the agent does not grow its egonet, it picks a partner among the members of its egonet. The probability that the agent $i$ picks the agent $j$ is proportional to the strength of the social affinity of $i$ towards $j$:
        $$P(i\xrightarrow{}j,t)=\frac{B_{i,j}(t)}{\sum_{k=1}^{N}B_{i,k}(t)}$$
    \end{enumerate}
    \setcounter{ADM}{\value{enumi}}
\end{enumerate}

$G(t)$ is given by the interactions determined by the previous steps, and the social bond interaction is then updated:
\begin{enumerate}
    \setcounter{enumi}{\value{ADM}}
    \item $B(t)$ is updated through a linear Hebbian process, meaning that for each edge $(i,j)$ in $G(t)$:
$$B_{i,j}(t+1)=B_{i,j}(t)+1$$
    \item $B(t+1)$ is pruned, taking possible loss of agents' memory into account. In the four models considered, this pruning consists in removing all directed bonds starting from $i$ with probability $p_{d}$ for each agent $i$. Once the pruning has been done, the previous steps are repeated for step $t+1$ instead of $t$.
\end{enumerate}

The only difference in rules between these three models is that triadic closure is allowed in ``ADM18conf16'', while in the two others, agents always grow their egonets by picking an unknown agent at random, i.e., $p_{u}=1$.
Parameters of each model are given in table \ref{tab:1}.

\subsection{Model ``ADM9conf16''}

This model differs from the ADM previously described on the following points:
\begin{itemize}
    \item the construction of $G(t)$
    \item the Hebbian process updating $B$
    \item the pruning process updating $B$
    \item the agent activity $a_{i}$ and the parameter $m_{i}$ (number of interactions emitted per agent) are agent-dependent.
    $a_{i}$ is drawn from a power-law of exponent $-1$ bounded between $a^{\text{min}}$ and $a^{\text{max}}$.
    $m_{i}$ is drawn from a uniform law on integers between 1 and $m^{\text{max}}$.
    These three parameters $a^{\text{min}}$, $a^{\text{max}}$ and $m^{\text{max}}$ are subjected to the genetic tuning evoked in the previous sub-section.
\end{itemize}

\subsubsection{Construction of $G(t)$}
Given a time step $t$, the network of interactions at that time $G(t)$ is built with two differences with respect to the ADM models from the previous sub-section:
\begin{itemize}
    \item taking into account a \textit{social context}:
    $G(t)$ is built both on the basis on $B(t)$ and $G(t-1)$.
    More precisely, the social affinity of $i$ towards $j$ is modulated by the number $c_{i,j}$ of their common partners at previous time:
    $$B_{i,j}(t)\xrightarrow{}\left(1+c_{i,j}(t-1)\right)B_{i,j}(t)$$
    \item taking into account \textit{contextual interactions}:
     we distinguish between \textit{intentional} and \textit{contextual} interactions.
    Intentional interactions are ``chosen'' by the agents; for the agent $i$, they consist in the $m_{i}$ interactions it emits.
    Contextual interactions are consequences of the fact that social interactions tend to be transitive.
    Hence, if $i$ is talking to $j$ who is talking to $k$, it is likely that $i$ and $k$ will also talk to each other.
    Thus contextual interactions can be viewed as a dynamic triadic closure mechanism.
    In our model, it is implemented as follows.
    Once the intentional interactions of every agent have been added in $G(t)$, we close every open triangle with some probability proportional to a new parameter $p_{c}$, which is subjected to genetic tuning \cite{le2023modeling}.
\end{itemize}

\subsubsection{Update of the social bond graph $B$}

Three points must be considered.

First, we distinguish between intentional and contextual interactions.
In the ``ADM9conf16'' model, we consider the latter as \textit{neutral}.
This means that contextual interactions do not lead to any change in the social ties:
they do not take part in the update of the social bond graph.

Second, social ties are reinforced through an exponential Hebbian process, instead of the linear process described before.
One important conceptual difference between the two processes is their sensitivity to the time ordering of past interactions:
only the number of past interactions matters in the linear Hebbian process, but the temporal order in which these interactions occur matters in the exponential Hebbian process.
This process is described in detail in \cite{gelardi2021temporal}.

Third, the pruning of $B$ consists in removing edges instead of nodes. This pruning is not uniform:
the stronger a social tie is, the less likely it is to be removed.
Moreover, ties that have been reinforced during the Hebbian process at time $t$ cannot be removed during the pruning at time $t$.
The probability of removing the directed edge $(i,j)$ from $B(t)$ is given by (see \cite{le2023modeling} for more details)
$$P_{d}(ij)=\exp\left(-\lambda {d_{i}^{\text{out}}} P(i\xrightarrow{}j)\right) \ , $$
where $d_{i}^{\text{out}}$ is the number of out-neighbours of $i$ in $B(t)$ and $P(i\xrightarrow{}j)$ is the probability that $i$ chooses to interact with $j$ at time $t$ (see previous sub-section).
$\lambda$ is a model parameter submitted to genetic tuning.

In table \ref{tab:1}, we give the values of each parameter of the four ADM models.

\begin{table*}
    \begin{tabular}{|c|p{3cm}|c|c|c|c|}
    \hline
        parameter notation & function & value in ``ADM18conf16'' & in ``min\_ADM1'' & in ``min\_ADM2'' & in ``ADM9conf16'' \\
        \hline
        $a$ & \centering probability of interaction proposal & 0.2 & 0.3 & 0.3 & $\in [0.03,0.91]$ \\
        \hline
        $m$ & \centering number of emitted interactions & 1 & 1 & 1 & $\in\llbracket1,2\rrbracket$ \\
        \hline
        $p_{g}$ & \centering rate of egonet growth & 0.004 & 0.085 & 0.085 & 0.102 \\
        \hline
        $1-p_{u}$ & \centering probability of triadic closure & 0.3 & 0 & 0 & 0.46 \\
        \hline
        $p_{d}$ & \centering rate of agents' memory loss & 0.02 & 0.02 & 0.1 & - \\
        \hline
        $\alpha$ & \centering evolution rate of social ties & - & - & - & 0.115 \\
        \hline
        $p_{c}$ & \centering dynamic triadic closure & - & - & - & 0.009 \\
        \hline
        $\lambda$ & \centering rate of edge pruning & - & - & - & 0.225 \\
        \hline
    \end{tabular}
    \caption{\label{tab:1}\textbf{Parameter values of the four ADM models considered in this paper.}
    The number of agents and the TN duration have been taken equal to the case of the empirical data set ``conf16''.
    For the models ``ADM9conf16'' and ``ADM18conf16'', the other parameters have been obtained by a genetic algorithm to match as closely as possible the TN ``conf16''.
    For the two other models, all parameters except the rate of memory loss $p_{d}$ have been set arbitrarily by hand.
    $p_{d}$ has been chosen according to the procedure described in Fig.~\ref{fig:14}.
    Contrary to the three other models, the $a$ and $m$ parameters are not the same for every agent in the model ``ADM9conf16''.
    The agent activity $a_{i}$ is drawn for each agent $i$ from a power-law of exponent $-1$ and bounded in the interval given in the table.
    The parameter $m_{i}$ is drawn for each agent $i$ from a uniform law on the set $\{1,2\}$.
    }
\end{table*}

\section{\label{sec:AppendixBB}EW models}
We introduced the EW (Edge-Weight) models to have a baseline.
Contrary to the ADM class, this class of models does not aim at reproducing empirical data.
They aim instead at answering the question:
How close from the empirical case are the flows of the almost simplest models we can build?

In the EW models, 
the number of nodes is fixed in time and denoted by $N$ like in the ADM models.
There is also a social bond graph $B$, indicating the strength of the social affinity between nodes.
Let us consider a time step $t$ and describe how the interaction graph $G(t)$ is built.
\begin{enumerate}
    \item  Each edge of non-zero weight in the social bond graph is activated at step $t$ with some probability.
To be activated at step $t$ means for an edge to belong to $G(t)$.
The activation probability is proportional to the edge weight, but the precise expression depends on the chosen rules for the TN dynamics:
\begin{enumerate}
\item  rule ``no shift'':
the probability that the edge $(i,j)$ is activated writes $\frac{B_{i,j}(t)}{t+1}$;
\item  rule ``with shift'':
the probability of activation writes $\frac{B_{i,j}(t)}{t+1-t_{b}(i,j)}$
where $t_{b}(i,j)$ is the time of last birth of the edge $(i,j)$.
The birth of an edge is defined by its entering into the social bond graph, i.e. the time at which $B_{i,j}$ becomes non-zero.
\end{enumerate}

\item  A constant number of edges selected at random among the edges of weight zero are activated.
\item  Weights of active edges are updated according to
$$B_{i,j}(t+1)=B_{i,j}(t)+1 \ .$$
\item  Before moving to step $t+1$, some edges are removed from the social bond graph, meaning their weights are set to zero.
In the EW class, two pruning processes are available:
\begin{enumerate}
    \item a node pruning, meaning that each node $i$ has the probability $p_{d}$ to be removed from $B(t+1)$; in practice, it means that all its edges are removed and it is replaced by an identical but isolated node;
\item  an edge pruning, meaning that each edge $(i,j)$ has the probability $p_{d}$ to be removed from $B(t+1)$.
\end{enumerate}

\end{enumerate}

Parameter values as well as the TN dynamics are described in table \ref{tab:2} for the three models considered in this paper:
min\_EW1, min\_EW2 and min\_EW3.

\begin{table*}
    \begin{tabular}{|c|c|c|c|c|}
    \hline
        rule name & associated parameter & value in min\_EW1 & in min\_EW2 & in min\_EW3 \\
        \hline
        shift & $t_{b}$ & 0 & last birth time & last birth time \\
        \hline
        birth rate & $n_{\text{new}}$ & 2 & 2 & 2 \\
        \hline
        pruning & $p_{d}$ & no pruning & node pruning, $p_{d}=0.01$ & edge pruning, $p_{d}=0.02$ \\
    \hline
    \end{tabular}
    \caption{\label{tab:2}\textbf{Parameter values of the three EW models considered in this paper.}
    Just like in the ADM models, the number of nodes and the TN duration were taken as equal to the case of the empirical data set ``conf16''.
    The value of $p_{d}$ has been chosen by hand, so that the number of temporal edges in the models match approximately the number of temporal edges in ``conf16''.
    }
\end{table*}

\newpage

\section{\label{sec:AppendixCC}Sizes of the considered data sets}

\begin{table*}[b]
    \begin{tabular}{|c|c|c|c|}
        \hline
        data set name & number of nodes & duration & number of temporal edges \\
        \hline
        ``conf16'' & 138 & 3 635 & 153 371 \\
        \hline
        ``conf17'' & 274 & 7 250 & 229 536 \\
        \hline
        ``conf18'' & 164 & 3 519 & 96 362 \\
        \hline
        ``conf19'' & 172 & 7 313 & 132 949 \\
        \hline
        ``utah'' & 630 & 1 250 & 353 708 \\
        \hline
        ``french'' & 242 & 3 100 & 125 773 \\
        \hline
        ``highschool1'' & 126 & 5 609 & 28 560 \\
        \hline
        ``highschool2'' & 180 & 11 273 & 45 047 \\
        \hline
        ``highschool3'' & 327 & 7 375 & 188 508 \\
        \hline
        ``hospital'' & 75 & 9 453 & 32 424 \\
        \hline
        ``malawi'' & 86 & 43 438 & 102 293 \\
        \hline
        ``baboon'' & 13 & 40 846 & 63 095 \\
        \hline
        ``work1'' & 92 & 7 104 & 9 827 \\
        \hline
        ``work2'' & 217 & 18 488 & 78 249 \\
        \hline \hline
        ``brownD01'' & 1 000 & 4 000 & 635 266 \\
        \hline
        ``brownD001'' & 988 & 4 000 & 661 878 \\
        \hline
        ``ABP2pi'' & 201 & 4 000 & 17 698 \\
        \hline
        ``ABPpi4'' & 1 000 & 4 000 & 156 173 \\
        \hline
        ``Vicsek2pi'' & 500 & 4 000 & 38 137 \\
        \hline
        ``Vicsekpi4'' & 500 & 4 000 & 41 841 \\
        \hline \hline
        ``ADM9conf16'' & 138 & 3 635 & 187 774 \\
        \hline
        ``ADM18conf16'' & 138 & 3 635 & 86 793 \\
        \hline
        ``min\_ADM1'' & 138 & 3 635 & 154 628 \\
        \hline
        ``min\_ADM2'' & 138 & 3 635 & 151 078 \\
        \hline
        ``min\_EW1'' & 138 & 3 635 & 57 376 \\
        \hline
        ``min\_EW2'' & 138 & 3 635 & 193 203 \\
        \hline
        ``min\_EW3'' & 138 & 3 635 & 185 901 \\
        \hline
    \end{tabular}
    \caption{\label{tab:3}\textbf{Sizes of the 27 data sets considered in this paper.}
    }
\end{table*}

\clearpage \newpage

\section{\label{sec:AppendixD}The motif\_error observable}

Given a temporal network, the motif\_error consists in comparing the observed and predicted probability of occurrence of motifs.
The observed probability of a motif is the ratio between its number of occurrences and the total number of motifs (nb\_tot) appearing in the considered temporal network.
Its predicted probability is instead computed under the hypothesis of statistical independence between its parts.
To simplify the discussion, we make the following hypotheses: (1) the motif we consider is an NCTN, (2) we denote its depth by $d$ and (3) we sample the motif on the untouched temporal network, i.e. $n=b=1$.

An NCTN is composed of one central node and $n_{sat}$ satellites, with $n_{sat}\geq1$.
Each satellite is characterized by an activity profile, which is a binary string of length equal $d$, with `1' indicating the satellite is interacting with the central node and `0' indicating it is not.
An NCTN is then given by the ordered collection $(A_{1},\ldots,A_{s})$ of the activity profiles of its $s$ satellites, where the order relation is the lexicographic order
(actually, any order other than the lexicographic order could be used, as long as it is a total order on the space of binary strings of length $d$).

The statistical independence hypothesis states that no correlation between satellites exist, so the occurrence probability of a motif is then the product of probabilities of the activity profiles of its satellites, as well as the probability of having the correct number of satellites.
This occurrence probability can thus be written as:
$$
P^{\text{th}}(A_{1},\ldots,A_{s})=P(n_{sat}=s)\prod_{i=1}^{s}P(A=A_{i})
$$
Note that $n_{sat}$ can be obtained from the instantaneous degree $k_{d}$ at aggregation level $d$, taking care that it cannot be zero:
$$P(n_{sat}=s)=P(k_{d}=s|k_{d}\geq1)=\frac{P(k_{d}=s)}{1-P(k_{d}=0)}$$
This makes possible the computation of $P(n_{sat}=s)$ without sampling the NCTN occurring in our network.
The activity profiles can also be sampled without explicitly computing the NCTN in the following way.
Activity profiles are collected from every node and every time of the temporal network.
Only the zero activity profile $\underbrace{0\ldots0}_{d}$ must be excluded, because empty motifs are not collected.
More precisely, for each time $t$ and each node $i$ such that $i$ is active at least once between times $t$ and $t+d-1$, we collect the activity profile of $i$ at $t$ of length $d$.
This profile is a binary string of `0' and `1'.
If positions along this string are indexed from 0 to $d-1$, a `1' in position $k$ means that the node $i$ is active at time $t+k$.
Then, the probability $P(A)$ is given by the fraction of occurrences $(i,t)$ that a node $i$ has $A$ as activity profile between from time $t$ to $t+d-1$.

Denoting the observed probability of occurrence of a motif by $P^{\text{obs}}$ and the set of motifs observed in our data set by $M$, we can now define the motif\_error as:
$$\sqrt{\frac{1}{|M|}\sum_{m\in M}\left(P^{\text{obs}}(m)-P^{\text{th}}(m)\right)^{2}} \ .$$

\section{\label{sec:AppendixF}Extracting the derivative sign with a neural network}

To extract the denoised sequence of the sign derivative of a given curve, we built and used a neural network.

\subsection{Architecture\label{subsec:1}}

The input layer is the rescaled data curve, where the rescaling of a curve $x_{i},i=1,\ldots,n$ consists in the following:
\begin{itemize}
    \item if the curve is constant, replace its value by 0.5;
    \item otherwise, replace $x_{i}$ by $\frac{x_{i}-m}{M-m}$, with $m=\min_{i}\{x_{i}\}$ and $M=\max_{i}\{x_{i}\}$.
\end{itemize}
Since each curve has 30 points at most in our case, the input layer has 30 neurons.
It can deal with curves with less points by repeating their last value until 30 points are filled.

Two output layers are used: one for the word size and one for its first letter.
Indeed, a word is fully determined by its first letter and its size.
This holds because curves we consider are either flat (so described by the word `0'), or they have no flat parts (so described by a word with no `0' in it).
The maximal word size we encountered in empirical cases was 3 so we assigned 3 neurons to the output coding for the word size.
As there are three possible letters, the output coding for the first letter has 3 neurons.
One-hot encoding was used for both outputs.

\begin{figure}
    \centering
    \includegraphics[width=0.7\columnwidth]{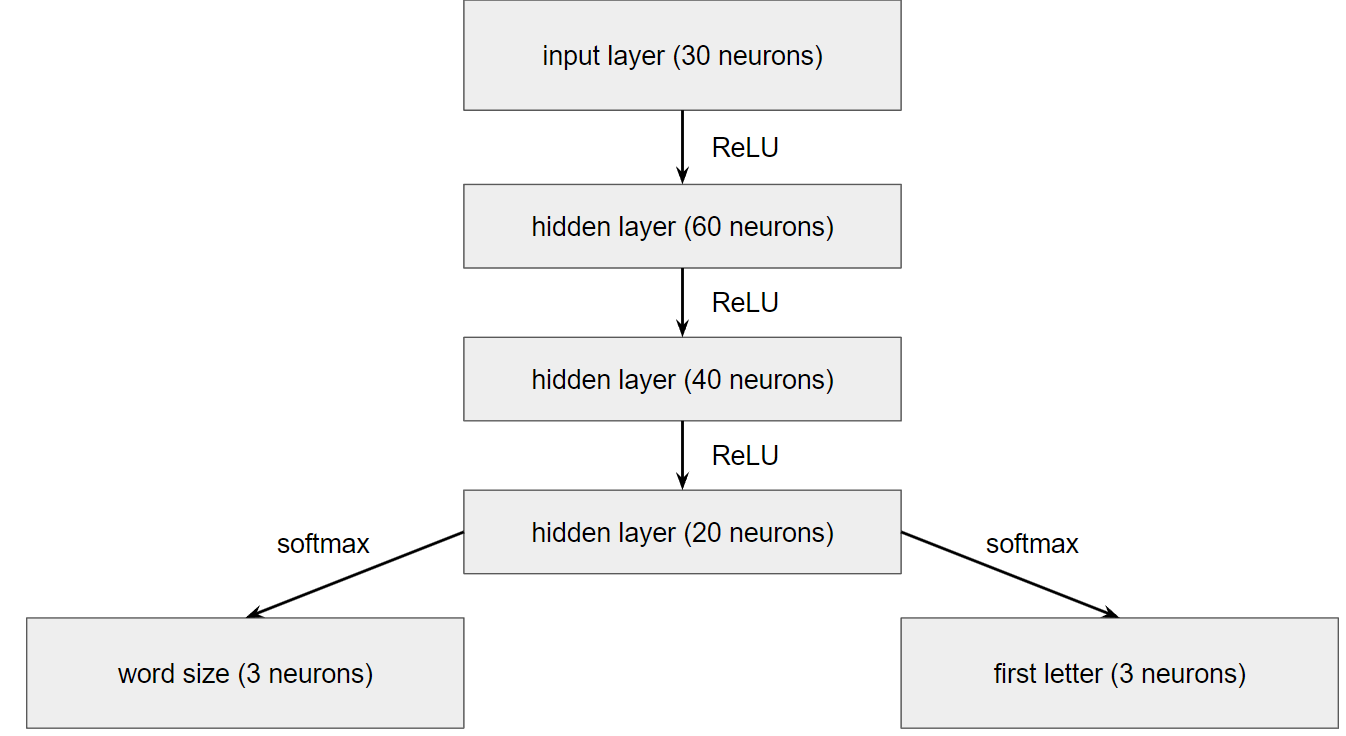}
    \caption{\label{fig:11}\textbf{Architecture of the neural network used for analog-symbolic conversion.}
    The input layer receives the rescaled and completed curve (see text).
    Since the word size is encoded by three neurons and one-hot encoding is used, this neural network can only produce words with a size either of 1, 2 or 3 letters.
    ReLU was used in every hidden layer as activation function, while softmax was used for the two outputs.
    The direction of the black arrows gives the information flow inside the network.
    }
\end{figure}

The overall architecture of the neural network is given in Fig.~\ref{fig:11}. Three intermediate layers were used.

\subsection{Training}
The training is supervised and the cost function is the categorical cross-entropy.
The number of epochs used for training has been chosen to maximize the validation accuracy, which was $91\%$ in our case.
Training set consists in 20\,000 examples of curves with various sizes, with known derivative sign sequence and endowed with white Gaussian noise as well as salt-and-pepper noise.
Salt-and-pepper noise consists in randomly choosing one point $1\leq i\leq n$ of the curve and replacing its value by either the curve maximum (probability 0.5) or the curve minimum (probability 0.5).

The training set is composed by curves described by various words.
Flat curves constitute $\frac{1}{7}\simeq14\,\%$ of the training set.
The remaining $\frac{6}{7}$ are equally split in three parts, each corresponding to a word size 1, 2 or 3.
Thus, the word `0' put apart, each size represents $\frac{2}{7}\simeq28\,\%$ of the training set.
For each size, examples with either letter `+' or `$-$' as first letter are present in equal proportion.
Here is the procedure to generate an example:
\begin{enumerate}
    \item choose a word size and first letter, as well as the curve length (number of points) randomly between 11 and 30 included;
    \item generate a curve of the chosen length whose derivative sign sequence matches the chosen word;
    \item rescale the curve according to the steps described in previous section \ref{subsec:1};
    \item add white Gaussian noise of standard deviation 0.05, then add salt-and-pepper noise with probability 0.05;
    \item rescale again the curve and repeat its last value to get 30 points:
    $$x_{1},\ldots,\underbrace{x_{n},\ldots,x_{n}}_{30-n+1\text{ copies}}\ .$$
\end{enumerate}

Let us detail the step 2, or how we generate a curve with a given sign sequence for its derivative.
Curves described by the word `0' are the simplest to generate:
we take $x_{i}=0.5,~\forall 1\leq i\leq n$.
Now let us consider a word of length $l$.
To build a curve described by this word, we first place the extrema of the curve and then fit a continuous function to fill the gaps between the extrema.
For a word of length $l$, we have $l+1$ points to place in the plane.
We denote those points by $(x_{0},y_{0}),\ldots,(x_{l},y_{l})$.
We put the first one of these points at the origin $(0,0)$ and the second one at the coordinates $(1,1)$ (resp. $(1,-1)$) if the first letter is `+' (resp. `$-$').
Said otherwise, introducing $\epsilon$ such that $\epsilon=0$ if the first letter is `$-$' and $\epsilon=1$ if the first letter is `+', we have:
$$
\begin{cases}
    x_{0}=0;~y_{0}=0\\
    x_{1}=1;~y_{1}=2\epsilon-1  \ .
\end{cases}
$$
The remaining points are placed one after the other, by pursuing to the right in a recursion process.

Note that we cannot place points at random, since we need to control two aspects of the curve, its maximum height ratio $r_{h}$ and its maximum slope ratio $r_{s}$. 
Height and slope are properties of segments joining two consecutive points $(x_{k},y_{k})$ and $(x_{k+1},y_{k+1})$, where $0\leq k\leq l-1$.
The height $h_{k}$ and the slope $s_{k}$ of such a segment are defined as:
$$
\begin{cases}
    h_{k}=|y_{k+1}-y_{k}|\\
    s_{k}=\frac{h_{k}}{x_{k+1}-x_{k}}
\end{cases}
$$
The maximum height and slope ratios are then defined as:
$$
\begin{cases}
    r_{h}=\frac{M_{h}}{m_{h}}\\
    r_{s}=\frac{M_{s}}{m_{s}}\\
    m_{h} = \min_{k}(h_{k});~M_{h} = \max_{k}(h_{k})\\
    m_{s} = \min_{k}(s_{k});~M_{s} = \max_{k}(s_{k})
\end{cases}
$$
To build our training set, we set $r_{h}=r_{s}=5$.
Ratios too high would lead to curves that seem discontinuous, and ratios too low would lead to curves that look flat, making impossible to guess the correct word.
A value of 5 seems to be a good compromise according to human eye appreciation.

Let us now describe how we generate our $l+1$ points recursively.
Let us assume $k$ points $(x_{0},y_{0}),\ldots,(x_{k-1},y_{k-1})$ have already been placed, $k\geq2$.
Let us denote the extrema height and slope computed at step $k-1$:
$$
\begin{cases}
    m_{h}^{(k-1)} = \underset{0\leq i\leq k-2}{\min}(h_{i});~M_{h}^{(k-1)} = \underset{0\leq i\leq k-2}{\max}(h_{i})\\
    m_{s}^{(k-1)} = \underset{0\leq i\leq k-2}{\min}(s_{i});~M_{s}^{(k-1)} = \underset{0\leq i\leq k-2}{\max}(s_{i})
\end{cases}
$$

We place the $(k+1)^{\text{th}}$ point at the right end of a segment line, starting from the last placed point.
Its height and slope must be compatible with the maximum ratios:
$$
\begin{cases}
    m_{h}^{(k-1)}r_{h}\leq h_{k-1}\leq\frac{M_{h}^{(k-1)}}{r_{h}}\\
    m_{s}^{(k-1)}r_{s}\leq s_{k-1}\leq\frac{M_{s}^{(k-1)}}{r_{s}}
\end{cases}
$$
To ensure this, we draw $h_{k-1}$ and $s_{k-1}$ uniformly at random between their bounds.
The recursion relation follows:
$$
\begin{cases}
    y_{k} = y_{k-1} + (-1)^{k+\epsilon}h_{k-1}\\
    x_{x} = x_{k-1} + \frac{h_{k-1}}{s_{k-1}}\\
    m_{h}^{(k)} = \min(m_{h}^{(k-1)},h_{k-1});~M_{h}^{(k)} = \max(M_{h}^{(k-1)},h_{k-1})\\
    m_{s}^{(k)} = \min(m_{s}^{(k-1)},s_{k-1});~M_{s}^{(k)} = \max(M_{s}^{(k-1)},s_{k-1})
\end{cases}
$$

Once our $l+1$ points $(x_{0},y_{0}),\ldots,(x_{l},y_{l})$ have been placed, we rescale both coordinates between 0 and 1.
For now, we only have the skeleton of our final curve ; it remains to dress it by adding intermediate points between the extrema.

\begin{figure}
    \centering
    \includegraphics[width=0.6\columnwidth]{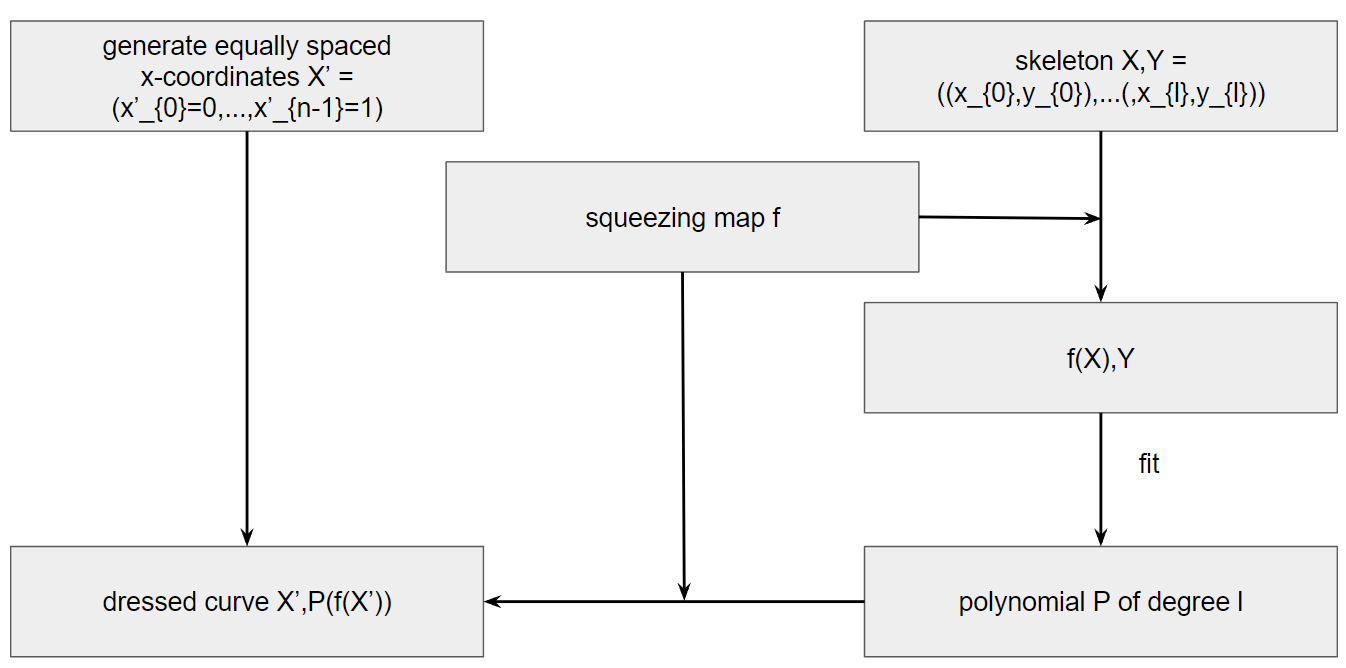}
    \caption{\label{fig:12}\textbf{Completion of a skeleton into one curve.}
    Once a dressed curve is obtained, we check that its maximum height and slope ratios do not exceed the specified bounds $r_{h}=r_{s}=5$.
    If so, a new dressing is performed with another squeezing map drawn at random among the 6 available, combined with a parameter $\gamma$ drawn at random too (see text).
    }
\end{figure}

The dressing principle is to fit a polynomial of degree $l$, so that the sign sequence of its derivative matches exactly the word we chose.
However, we squeeze our skeleton before fitting to ensure diversity in the examples we feed to the neural network:
squeezing is done by applying a map to the $x$-coordinates, chosen among 6 different maps.
The dressing procedure is fully described in Fig. \ref{fig:12}.

The 6 maps we use are:
\begin{itemize}
    \item `raw': $x\mapsto x$
    \item `inv': $x\mapsto \frac{1}{1+\gamma x}$
    \item `exp': $x\mapsto \exp(\gamma x)$
    \item `expinv': $x\mapsto \exp(-\gamma x)$
    \item `log': $x\mapsto\log(1+\gamma x)$
    \item `tanh': $x\mapsto\tanh(\gamma x)$
\end{itemize}
where $\gamma$ is a float parameter drawn at random between 1 and 10  after a map is chosen.

Once the skeleton has been dressed into a curve of $n$ points, the final curve consists in the rescaled sequence of the $y$-coordinates $y_{0},\ldots,y_{n-1}$.

\begin{figure}
    \includegraphics[width=0.4\columnwidth]{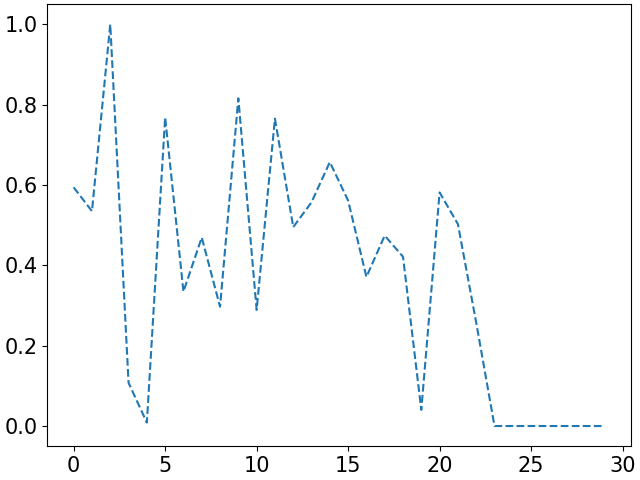}
    \includegraphics[width=0.4\columnwidth]{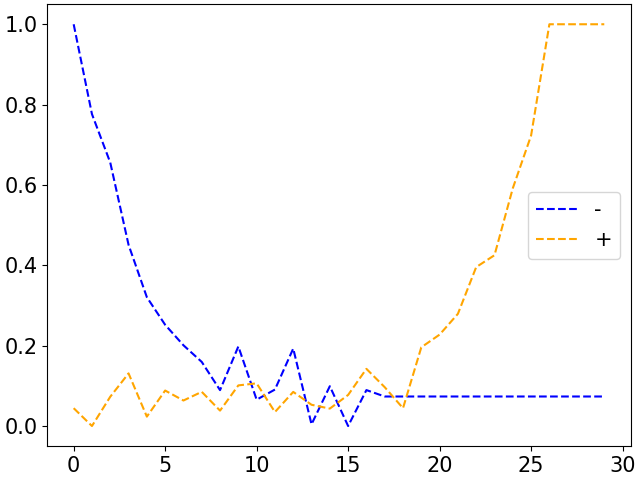}
    \includegraphics[width=0.4\columnwidth]{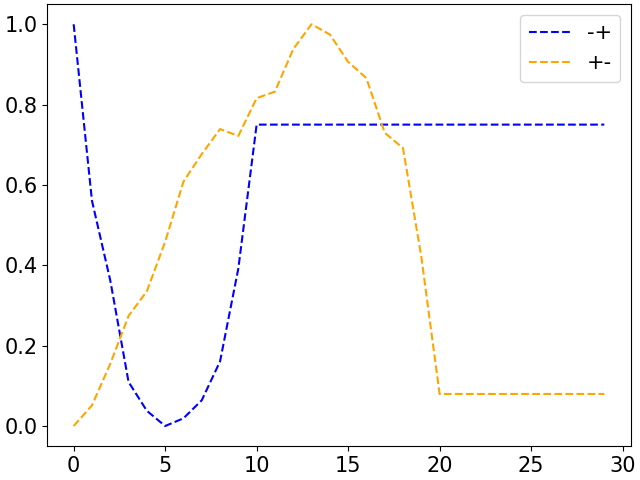}
    \includegraphics[width=0.4\columnwidth]{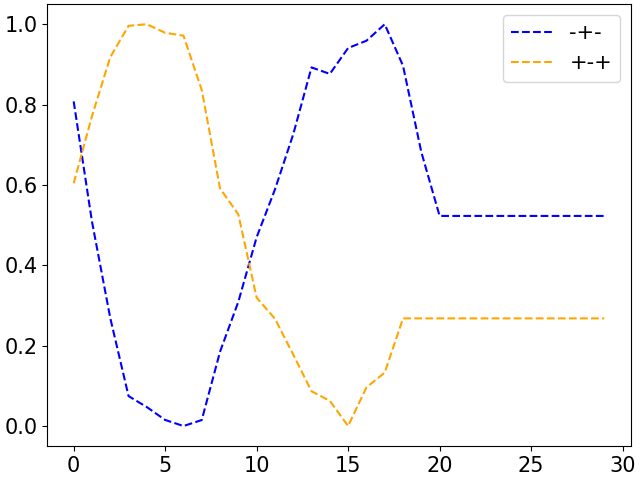}
    \caption{\label{fig:13}\textbf{Examples of curves used for training the neural network.}
    The displayed curves have been drawn at random from the training set of 20 000 labelled examples fed to the neural network for its supervised training.
    We drew one example per word present in the training set.
    The legend indicates the training labels of the curves.
    The curve shown at the top left is labelled by the word `0'.
    }
\end{figure}

Examples of final curves are displayed in Fig. \ref{fig:13} together with their associated words.

\section{Nomenclature of observables\label{sec:app_nomenclature}}

In section \ref{subsec:3}, we described various observables of a temporal network.
Here, we explain how we name an observable in a standardized way, which is suitable in particular for numerical analysis.

Recall that we introduced two main types of observables:
distribution and motif-based observables.
We name distribution observables by concatenation of a prefix and a suffix separated by a `0'.
The prefix indicates the observed object and the suffix gives the object property we care about.
In this paper, an object is a node or an edge while a property is e.g. its duration, time\_weight, etc (cf \ref{subsubsec:distr_obs}).
Each distribution observable gives rise to two scalar ones by following the naming convention:
$$\text{object}0\text{property}0\text{scalar\_type}$$
We considered only two scalar types:
``avg'' for $\langle x\rangle$ and ``frac'' for $\frac{\langle x^{2} \rangle}{\langle x\rangle}$. For example, ``node0duration0avg'' refers to the average number of consecutive time steps a node is active.

The naming convention of motif-based observables is simpler:
$$\text{object}0\text{property}$$
However, an object is given by a motif type and its depth $d$ together, written as:
$$d\text{-motif\_type}$$
A property is then any item from the list of sub-subsection \ref{subsubsec:3}. For example, 2-NCTN0nb\_tot is the total number of NCTN of depth 2, and 3-ECTN0sim\_trunc0vs\_b is the degree of invariance of the ECTN vector of depth 3 under STS.
At level $n$, this invariance is quantified by the cosine similarity between the vectors obtained for TN$(n,1)$ and TN$(n,b)$.

Note that, when computing this similarity, we must precise whether we compute it versus $b$ or versus $n$.
Thus, we distinguish between internal and external observables, because we cannot use the same algorithm to compute them:
\textit{Internal} observables can be computed for every couple $(n,b)$ using only the transformed data set TN$(n,b)$.
They are the majority of the observables considered here, like e.g. the total number of motifs or the average node duration.
\textit{External} observables need some additional data to be computed.
For example, the similarity between TN$(n,b)$ and TN$(1,b)$ is an external observable because it requires TN$(1,b)$ besides TN$(n,b)$ to be computed.

\begin{figure}
    \centering
    \includegraphics[width=0.6\columnwidth]{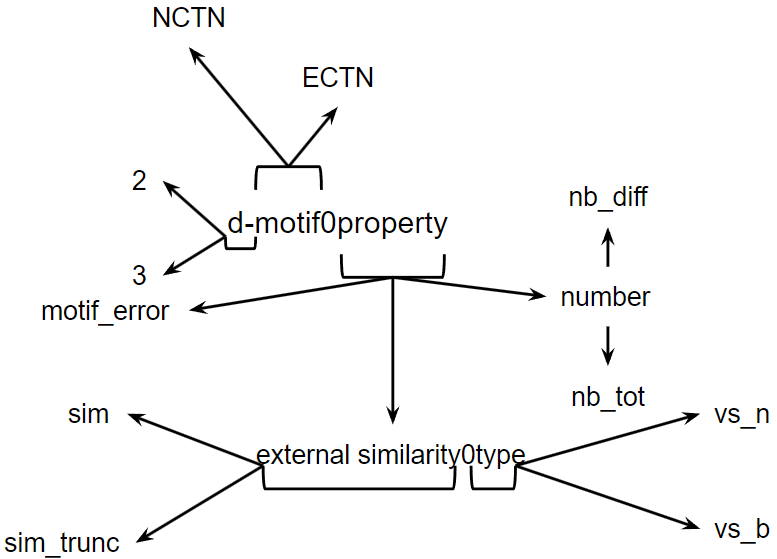}
\caption{\label{fig:7}\textbf{Nomenclature of motif-based observables.}
All possible names have been encoded in a tree.
A name is read by choosing recursively a child for each parent until leaves are reached.
Going from a parent to a child is done by following the arrows.
Here are examples of names generated according to this principle:
``3-ECTN0sim\_trunc0vs\_b'', ``2-NCTN0nb\_tot''.
}
\end{figure}

\begin{figure*}
    \includegraphics[width=\textwidth]{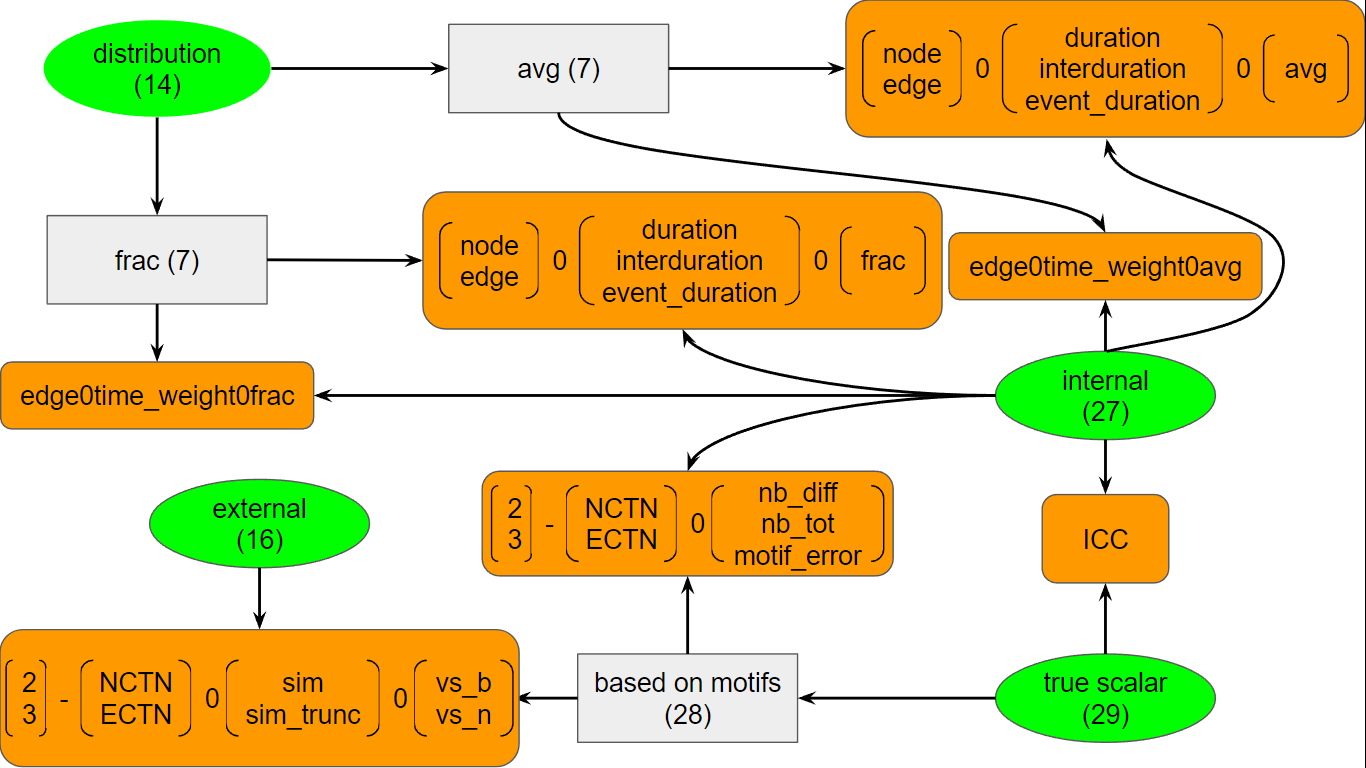}
    \caption{\label{fig:15}\textbf{Recap of the types of observables considered in this paper.}
    Observables have been displayed by using a directed acyclic graph.
    Leaves are the rounded boxes colored in orange. They contain observables of the same type.
    In the non-leaves boxes are indicated a type and the number of observables of this type among the ones considered in this paper.
    Roots are highlighted as elliptic boxes colored in green.
    The gray rectangular boxes correspond to the levels intermediate between roots and leaves.
    An arrow from a box $i$ to a box $j$ means that $i$ contains $j$.
    For example, the observable edge0time\_weight0avg is both an internal and a scalarized observable of type avg.
    ``True scalar'' observables are just the observables that are not considered as deriving from a distribution.
    Note that some boxes contain column vectors.
    It means that these boxes contain every observable that is possible to write by choosing any component for each column vector.
    For example, the out-neighbour of the box ``frac (7)'' contains $2\times3\times1=6$ observables.
    }
\end{figure*}

On Figure \ref{fig:7}, we propose a diagrammatic representation of the nomenclature of all motifs-based observables. On Figure \ref{fig:15}, we summarize every type of observable encountered in this paper, and how they relate to each other.

\end{document}